\documentclass[a4paper,usenatbib]{mnras}

\usepackage{url}
\usepackage{deluxetable}
\usepackage{newtxtext,newtxmath}
\usepackage{ae,aecompl}
\usepackage{amssymb, amsmath}
\usepackage{graphicx}
\usepackage{natbib}
\usepackage{hyperref}
\usepackage{cleveref}
\usepackage[usenames, dvipsnames]{color}
\usepackage{colortbl}
\usepackage{fontawesome}

\hypersetup{colorlinks=true,
            citecolor=MidnightBlue,
            linkcolor=MidnightBlue,
            filecolor=magenta,      
            urlcolor=cyan}

\DeclareGraphicsExtensions{.pdf,.png,.jpg}

\defcitealias{Huang2018b}{Paper~I}	
\defcitealias{Huang2018c}{Paper~II}	
    
\def\arcsec{{\prime\prime}}

\def\aj{{AJ}}

\def\apj{{ApJ}}
\def\apjs{{ApJS}}
\def\asec{$^{''}$}

\def\lax{{$\mathrel{\hbox{\rlap{\hbox{\lower4pt\hbox{${\sim}$}}}\hbox{$<$}}}$}}
\def\gax{{$\mathrel{\hbox{\rlap{\hbox{\lower4pt\hbox{${\sim}$}}}\hbox{$>$}}}$}}
\def\simlt{\lower.5ex\hbox{$\; \buildrel < \over {\sim} \;$}}
\def\simgt{\lower.5ex\hbox{$\; \buildrel > \over {\sim} \;$}}

\def\mnras{{MNRAS}}
\def\nat{{Nature}}
\def\pasp{{PASP}}

\def\etal{{\ et al.}}

\def\cmodel{\texttt{cModel}}

\def\logmh{{$\log_{10} (M_{\mathrm{vir}}/M_{\odot})$}}
\def\logmvir{{$\log_{10} M_{\mathrm{vir}}$}}
\def\logmpeak{{$\log_{10} M_{\mathrm{peak}}$}}

\def\mtot{{$M_{\star}^{\mathrm{Tot}}$}}

\def\logminn{{$\log_{10} (M_{\star,10\mathrm{kpc}}/M_{\odot})$}}
\def\logmtot{{$\log_{10} (M_{\star,100\mathrm{kpc}}/M_{\odot})$}}

\def\logmmax{{$\log_{10} (M_{\star,\mathrm{max}}/M_{\odot})$}}

\def\logmcmod{{$\log_{10} (M_{\star,\mathrm{cmodel}}/M_{\odot})$}}

\def\logm10{{$\log (M_{\star,10\ \mathrm{kpc}}/M_{\odot})$}}
\def\logm30{{$\log (M_{\star,30\ \mathrm{kpc}}/M_{\odot})$}}
\def\logm50{{$\log (M_{\star,50\ \mathrm{kpc}}/M_{\odot})$}}
\def\logm100{{$\log (M_{\star,100\ \mathrm{kpc}}/M_{\odot})$}}
\def\logmtot{{$\log (M_{\star,\mathrm{tot}}/M_{\odot})$}}

\def\m2l{{$M_{\star}/L_{\star}$}}
\def\s2n{{$\mathrm{S}/\mathrm{N}$}}
\def\mden{{$\mu_{\star}$}}

\def\insitu{{\textit{in situ}}}
\def\exsitu{{\textit{ex situ}}}
\def\ins{{\textit{in situ}}}
\def\exs{{\textit{ex situ}}}

\def\mvir{{$M_{\mathrm{vir}}$}}
\def\mhalo{{$M_{\mathrm{vir}}$}}
\def\mpeak{{$M_{\mathrm{peak}}$}}
\def\mhost{{$M_{\mathrm{200b}}$}}
\def\mh200b{{$M_{\mathrm{200b}}$}}
\def\mh200c{{$M_{\mathrm{200c}}$}}

\def\mstar{{$M_{\star}$}}

\def\mall{{$M_{\star}^{\mathrm{all}}$}}
\def\mcen{{$M_{\star}^{\mathrm{cen}}$}}
\def\mgal{{$M_{\star}^{\mathrm{gal}}$}}
\def\mins{{$M_{\star}^{\mathrm{ins}}$}}
\def\mexs{{$M_{\star}^{\mathrm{exs}}$}}

\def\fgal{{$\delta_{\rm gal}$}}

\def\fins{{$\delta_{\rm ins}$}}
\def\fexs{{$\delta_{\rm exs}$}}

\def\minn{{$M_{\star}^{10}$}}
\def\mtot{{$M_{\star}^{100}$}}
\def\mmax{{$M_{\star}^{\mathrm{max}}$}}
\def\mcmodel{{$M_{\star}^{\mathrm{cmod}}$}}

\def\mallmod{{$\mathcal{M}_{\star}^{\mathrm{all}}$}}
\def\mgalmod{{$\mathcal{M}_{\star}^{\mathrm{gal}}$}}
\def\mcenmod{{$\mathcal{M}_{\star}^{\mathrm{cen}}$}}
\def\minsmod{{$\mathcal{M}_{\star}^{\mathrm{ins}}$}}
\def\mexsmod{{$\mathcal{M}_{\star}^{\mathrm{exs}}$}}

\def\minsten{{$\mathcal{M}_{\star}^{\mathrm{in, 10}}$}}
\def\mexsten{{$\mathcal{M}_{\star}^{\mathrm{ex, 10}}$}}
\def\minnmod{{$\mathcal{M}_{\star}^{10}$}}
\def\mmaxmod{{$\mathcal{M}_{\star}^{\mathrm{Max}}$}}

\def\um{\texttt{UniverseMachine}}
\def\htools{\texttt{halotools}}
\def\smdpl{\texttt{SMDPL}}
\def\rockstar{\texttt{Rockstar}}
\def\emcee{\texttt{emcee}}
\def\asap{\texttt{ASAP}}
\def\redm{\texttt{redMaPPer}}

\def\dsigma{{$\Delta\Sigma$}}

\title[Halo mass and stellar mass profiles]{Weak Lensing Reveals a Tight Connection 
Between Dark Matter Halo Mass and the Distribution of Stellar Mass in Massive Galaxies}
       
\author[S. Huang et al.]{
        Song Huang$^{1,2}$\thanks{E-mail: shuang89@ucsc.edu (SH)},
        Alexie Leauthaud$^{1}$,
        Andrew Hearin$^{3}$,
        Peter Behroozi$^{4}$,
        \newauthor
        Christopher Bradshaw$^{1}$,
        Felipe Ardila$^{1}$,
        Joshua Speagle$^{5}$,
        Ananth Tenneti$^{6}$,
        \newauthor
        Kevin Bundy$^{7}$,
        Jenny Greene$^{8}$,
        Crist{\'o}bal Sif{\'o}n$^{8}$,
        Neta Bahcall$^{8}$
        \\
        $^{1}$Department of Astronomy and Astrophysics, University of California 
              Santa Cruz, 1156 High St., Santa Cruz, CA 95064, USA\\
        $^{2}$Kavli-IPMU, The University of Tokyo Institutes for Advanced Study, 
              the University of Tokyo (Kavli IPMU, WPI), \\ ~~Kashiwa 277--8583, Japan\\  
        $^{3}$Argonne National Laboratory, Argonne, IL 60439, USA\\
        $^{4}$Department of Astronomy and Steward Observatory, University of Arizona, 
              Tucson, AZ 85721, USA\\
        $^{5}$Department of Astronomy, Harvard University, 60 Garden St, MS 46, 
              Cambridge, MA 02138, USA\\ 
        $^{6}$McWilliams Center for Cosmology, Department of Physics, 
              Carnegie Mellon University, Pittsburgh, PA 15213, USA \\          
        $^{7}$UCO/Lick Observatory, University of California, Santa Cruz,
              1156 High Street, Santa Cruz, CA 95064, USA\\
        $^{8}$Department of Astrophysical Sciences, Peyton Hall,
              Princeton University, Princeton, NJ 08540, USA
        } 
          
\date{Accepted XXX. Received YYY; in original form ZZZ}        
\pubyear{2018}                                  
  
\begin{document}

\label{firstpage}
\pagerange{\pageref{firstpage}--\pageref{lastpage}}

\maketitle


\begin{abstract}
      
    Using deep images from the Hyper Suprime-Cam (HSC) survey and taking advantage of 
    its unprecedented weak lensing capabilities, we reveal a remarkably tight connection 
    between the stellar mass distribution of massive central galaxies and their host 
    dark matter halo mass.
    Massive galaxies with more extended stellar mass distributions tend to live in 
    more massive dark matter haloes. 
    We explain this connection with a phenomenological model that assumes,  
    (1) a tight relation between the halo mass and the total stellar content in the halo,
    (2) that the fraction of \ins{} and \exs{} mass at $r<$10 kpc depends on halo mass. 
    This model provides an excellent description of the stellar mass functions (SMF)
    of total stellar mass (\mmax{}) and stellar mass within inner 10 kpc (\minn{}) 
    and also reproduces the HSC weak lensing signals of massive galaxies with different 
    stellar mass distributions. 
    The best-fit model shows that halo mass varies significantly at fixed total stellar
    mass (as much as 0.4 dex) with a clear dependence on \minn{}.
    Our two-parameter \mmax{}--\minn{} description provides a more accurate picture 
    of the galaxy--halo connection at the high-mass end than the simple stellar--halo 
    mass relation (SHMR) and opens a new window to connect the assembly history of 
    halos with those of central galaxies.  
    The model also predicts that the \exs{} component dominates the mass profiles 
    of galaxies at $r< 10$ kpc for $\log M_{\star} \ge 11.7$). 
    The code used for this paper is available online 
    \href{https://github.com/dr-guangtou/asap}{\faGithubAlt} 
        
\end{abstract}




\section{Introduction}
    \label{sec:intro}
    
    
    During the last decade, observations and hydrodynamic simulations have 
    significantly furthered our understanding of the formation of massive galaxies.
    The observed gradual mass assembly (e.g. \citealt{Lundgren2014, Ownsworth2014, 
    Bundy2017}) and dramatic structural evolution (e.g. \citealt{vdWel2014, 
    Clauwens2017, Hill2017}) of massive galaxies support a `two-phase' scenario
    for their formation. 
    (e.g. \citealt{Oser2010, Oser2012, RodriguezGomez2016}). 
    Under this picture, intense dissipative processes at high-redshift swiftly 
    builds up the massive, compact `core' of today's massive galaxies
    (e.g. \citealt{Damjanov2009, Toft2014, vanDokkum2015}), including most of the 
    \ins{} component: stars formed in the main progenitor of the host dark matter 
    halo (e.g. \citealt{DeLucia2007, Genel2009}). 
    Super massive galaxies, however, are also expected to have a large \exs{} 
    component: stars that are accreted from other haloes. 
    After the quenching of star formation in massive galaxies,
    (e.g. \citealt{Hopkins2008b, Johansson2009, Conroy2015}), the gradual 
    accumulation of the \exs{} component dominates the assembly of massive galaxies 
    and helps build up extended stellar envelopes
    (e.g. \citealt{vanDokkum2008, Bezanson2009, Patel2013, Huang2013b}). 
    More importantly, these two components should show differences in their spatial 
    distributions as a large fraction of the \exs{} component is expected to be 
    deposited at large radii (e.g. \citealt{Hilz2013, Oogi2013}). 
    This suggests that the stellar mass distribution of massive galaxies contains 
    information about their assembly history.
            
    
    From a cosmological perspective, to understand the assembly of massive 
    galaxies is to understand how they hierarchically grow with their dark matter 
    haloes (e.g. \citealt{Wechsler2018} and the references within). 
    Recently, the basic understanding of the stellar--halo mass relation (SHMR) 
    has been established using various direct and indirect methods
    (e.g. \citealt{Hoekstra2007, More2011, Leauthaud2012, Behroozi2013, Coupon2015,
    Zu2015, vanUitert2016, Tinker2017, Shan2017, Kravtsov2018}).
    At low redshift, the SHMR can be characterized by a a power-law relation at low 
    masses, a characteristic pivot halo mass, and an exponential rise at higher 
    masses (\citealt{Behroozi2013, RodriguezPuebla2017, Moster2018}). 
    Constraints on the SHMR have helped us gain insight into the 
    galaxy--halo connection, but an in-depth picture about how the assembly of 
    galaxies is tied to their dark matter haloes is still lacking.
    At high-mass end, the situation is particularly true (e.g. \citealt{Tinker2017, 
    Kravtsov2018}).
    First, challenges in measuring the total stellar mass of massive elliptical 
    galaxies with extremely extended light profile (e.g. \citealt{Bernardi2013, 
    Bernardi2014, Bernardi2017, Huang2018b, Kravtsov2018}; 
    \citealt{Pillepich2017b}) 
    complicate constraints of the SHMR for massive galaxies.
    More importantly, this simple scaling relation does not provide the full picture; 
    specifically, it does not describe whether or not the internal structure 
    (i.e., the way in which stellar mass is distributed in massive galaxies) is 
    tied to the assembly history of their dark matter haloes.
    At high stellar mass (\mstar{}) end, the scatter of halo mass at fixed stellar 
    mass is of order 0.3--0.4 dex \citep[e.g.,][]{Tinker2017}). 
    In this paper, we seek to explain how similarly massive galaxies can live in 
    haloes with very different mass and assembly histories, by looking for 
    signatures of this assembly process in the stellar mass profiles of massive 
    galaxies. 
    
	 
	In previous work (\citealt{Huang2018b}, \citealt{Huang2018c}; 
	\citetalias{Huang2018b} and \citetalias{Huang2018c} hereafter), we map the stellar
	mass distributions of massive galaxies at $0.3 \le z < 0.5$ to $>100$ kpc
	individually using deep images from the 
	\href{https://www.naoj.org/Projects/HSC/}{Hyper Suprime-Cam}
	(HSC; \citealt{Miyazaki2012}) 
	\href{https://hsc.mtk.nao.ac.jp/ssp/}{Subaru Strategic Program}
	(SSP, hereafter `HSC survey'; \citealt{HSC-SSP, 
	HSC-DR1}). 
	With the help of deep images and the 
	\href{http://risa.stanford.edu/redmapper/}{\redm{} cluster catalog} (e.g. 
	\citealt{Rykoff2014, Rozo2014}),
	we find evidence that the surface stellar mass density profiles (\mden{}) of massive 
	central galaxies depend on dark matter halo mass: centrals galaxies in more massive 
	halos tend to have more extended stellar mass distributions (also see: 
	\citealt{Charlton2017, Yoon2017}) and less mass in the inner 10 kpc (\minn{}).
	
	Here we seek to directly confirm this dependence and characterize this relation 
	using the galaxy--galaxy weak lensing (`g--g lensing') method 
	(e.g. \citealt{Mandelbaum2006a, Mandelbaum2006b, Leauthaud2012, Coupon2015,
	Leauthaud2017}) that probes the dark matter halo mass distribution by  
	measuring the coherent shape distortion of background galaxies.
	Instead of relying on a cluster catalog, the unprecedented g--g lensing capability 
	of the HSC survey (e.g. \citealt{HSC-WLCAT, Medezinski2018, Miyatake2018}) allows 
	us to map the halo mass trend across a 2--D plane described 
	by the \minn{} and stellar mass within the largest aperture that is allowed by 
	the depth of the image (\mmax{}) and build an empirical model for galaxy--halo 
	connection at high-mass end.

       
    This paper is organized as follows. 
    We briefly summarize the sample selection and data reduction processes in 
    \S \ref{sec:data}.  
    Please refer to \citetalias{Huang2018b} for more technical details.
    \S \ref{sec:dsigma} describes the weak lensing methodology, and the 
    measurements of aperture \mstar{} and \mden{} profiles are discussed in 
    \S \ref{sec:measure}.
    In \S \ref{sec:model}, we introduce an empirical model to describe the relation
    between dark matter halo mass and the distribution of stellar mass within super 
    massive galaxies. 
    The results from our best-fit model are presented in \S \ref{sec:result} and 
    discussed in \S \ref{sec:discussion}. 
    Our summary and conclusions are presented in \S \ref{sec:summary}.

    
    All magnitudes in this work are in AB system  (\citealt{Oke1983}) and 
    have been corrected for Galactic extinction using \citet{Schlafly11}.
    For cosmology, we assume $H_0$ = 70~km~s$^{-1}$ Mpc$^{-1}$, 
    ${\Omega}_{\rm m}=0.3$, and ${\Omega}_{\rm \Lambda}=0.7$.
    Stellar mass (\mstar{}) is derived using a Chabrier initial mass function 
    (IMF; \citealt{Chabrier2003}). 
    And we use the virial mass for dark matter halo mass (\mhalo{}) as 
    defined in \citealt{BryanNorman1998}.
    


\section{Data and Sample Selection}
    \label{sec:data}

\subsection{SSP S16A data}
    \label{sec:s16a}      
    
	
	
	In this work, we use the \texttt{WIDE} layer of the internal data release 
	\texttt{S16A} of the HSC SSP, an ambitious imaging survey using the new prime 
	focus camera on the 8.2-m Subaru telescope. 
	These data are reduced by \texttt{hscPipe 4.0.2}, a specially tailored version of 
	\href{https://pipelines.lsst.io}{the Large Synoptic Survey Telescope (LSST) pipeline} 
	(e.g.\ \citealt{Juric2015}; \citealt{Axelrod2010}), 
	modified for HSC (\citealt{HSC-PIPE}).
	The coadd images are $\sim$3--4 mag deeper than SDSS (Sloan Digital Sky Survey; 
	e.g. \citealt{SDSS-DR7, SDSS-DR8, SDSS-DR12}), with a pixel scale of 
	0$^{\arcsec}.168$. 
	The seeing in the $i$-band has a mean full width at half maximum (FWHM) of 
	0$^{\arcsec}.58$.
	Please refer to \citet{HSC-SSP, HSC-DR1} for more details about the survey 
	design and the data products.
	The general performance of \texttt{hscPipe} is validated using a synthetic object 
	pipeline \texttt{synpipe} (e.g. \citealt{SynPipe}; code available on GitHub at this link
	\href{https://github.com/lsst/synpipe}{\faGithubAlt}).
	In addition to the full-color and full-depth cuts, regions that are affected 
	by bright stars are also masked out \citet{HSC-STAR}.
    
    
    The HSC collaboration compiles the spectroscopic redshifts (spec--$z$ hereafter) 
    of HSC galaxies from a series of available spectroscopic surveys, which is 
    the main sources of spec-$z$ in this work.
    We also include additional spec-$z$ from the most recent data release of 
    the \href{http://www.gama-survey.org/dr3}{Galaxy And Mass Assembly} (GAMA) survey
    \citep{Driver2009, Driver2011, Liske2015, Baldry2018} which 
    significantly overlaps with HSC coverage in their G02, G09, G12, and G15 regions
    and greatly improve the spec-$z$ completeness of our massive galaxy sample. 
    The HSC collaboration also provides photometric redshift (photo-$z$ hereafter) 
    measurements using the point spread function (PSF)--matched five-band fluxes 
    within 1$^{\arcsec}.5$ apertures and six different algorithms. 
    Here, we use the spec-$z$ sample and the photo-$z$ measurements based on 
    \href{https://github.com/joshspeagle/frankenz}{the Flexible Regression over 
    Associated Neighbors with Kernel dEnsity 
    estimatioN for Redshifts} (\texttt{FRANKEN-Z}; Speagle\etal{} in prep.) algorithm.
    Please refer to \citet{HSC-PHOTOZ} for details about photo--$z$ catalogues. 
    
    
    For our weak lensing measurements, we make use of the first-year shear catalogue
    described in detail by \citet{HSC-WLCAT}. 
    Currently, we use the re-Gaussianization algorithm (\citealt{HirataSeljak2003}) 
    to measure galaxy shapes on $i$-band coadd images.
    Please see \citet{HSC-WLCAT} and \citet{HSC-WLCALIB} for more details about our
    shape measurements and their calibration.
    Our shape catalogue also excludes a small fraction of the survey area that has a 
    problematic PSF model. 
    The resulting survey area is the full--depth full--color region for weak lensing 
    analysis (\texttt{WLFDFC}) region, which covers ${\sim} 137$ deg$^2$ in all five 
    bands ($grizy$) to the required imaging depth (5$\sigma$ point source detection 
    limit of 26.0 mag). 
    For our g--g lensing measurements, we also use a random catalogue that contains a 
    half million objects and covers the \texttt{WLFDFC} area 
    (e.g. \citealt{Singh2017, HSC-STAR}).

\subsection{Sample selection}
    \label{sec:sample}   

    
    
    Our sample selection is very similar to \citet{Huang2018b} and \citet{Huang2018c} 
    (\citetalias{Huang2018b} and \citetalias{Huang2018c} hereafter).
    We select all galaxies with $i_{\rm CModel} <= 22.0$ mag and useful five-band 
    \cmodel{} photometry in the \texttt{WLFDFC} area.
    Instead of only using galaxies with spec--$z$'s however, we now assign a best 
    redshift ($z_{\rm best}$) to each object: 
    We adopt the spec--$z$ when it is available; for others, we use the photo--$z$
    measurements from \texttt{FRANKEN-Z} as $z_{\rm best}$.
    We select all galaxies within $0.19 < z_{\rm best} < 0.51$, where redshift 
    evolution is not a serious concern and the volume is large enough 
    ($1.03\times 10^8$ Mpc$^3$) to ensure a large sample of massive galaxies.
    The performance of \texttt{FRANKEN-Z} at this redshift and magnitude range is 
    unbiased and reliable with respect to the training sample. 
    The typical $1\sigma$ uncertainty is $\sim7$ per cent with a median bias of
    about $-0.3$ per cent and typical outlier fraction of 11 to 19 per cent in this
    redshift range.
    Compared with the spec-$z$--only sample, adding in the photo-$z$'s greatly 
    improves the \mstar{} completeness of our sample but does not alter any of 
    our key results.
    
    
    We perform five-band spectral energy distribution (SED) fitting using the \cmodel{} 
    photometry to derive the average mass--to--light ratio (\m2l{}) of galaxies and initial 
    estimates of \mstar{} (\mcmodel{}).
    The SED fitting procedure is identical to the one used in \citetalias{Huang2018b}. 
    In short, we use 
    \href{http://www.sos.siena.edu/~jmoustakas/isedfit/}{\texttt{iSEDFit}}
    (\citealt{Moustakas2013}) to measure \m2l{} ratios and $k$-corrections, assuming 
    the \citet{Chabrier2003} IMF and using the \href{https://github.com/cconroy20/fsps}{
    Flexible Stellar Population Synthesis models}
    (FSPS; \texttt{v2.4}; \citealt{FSPS}, \citealt{Conroy2010}).
    Please refer to \citetalias{Huang2018b} for more details. 
    Based on the SED fitting results, we select galaxies with \logmcmod{}$>10.8$ as the 
    initial sample of massive galaxies. 
    Typical uncertainty of \mcmodel{} is around 0.05 to 0.1 dex.
    We further measure the \mden{} profiles of these galaxies and aperture \mstar{}
    within different radii (see \ref{ssec:mprof}).

  \begin{figure*}
      \centering
      \includegraphics[width=\textwidth]{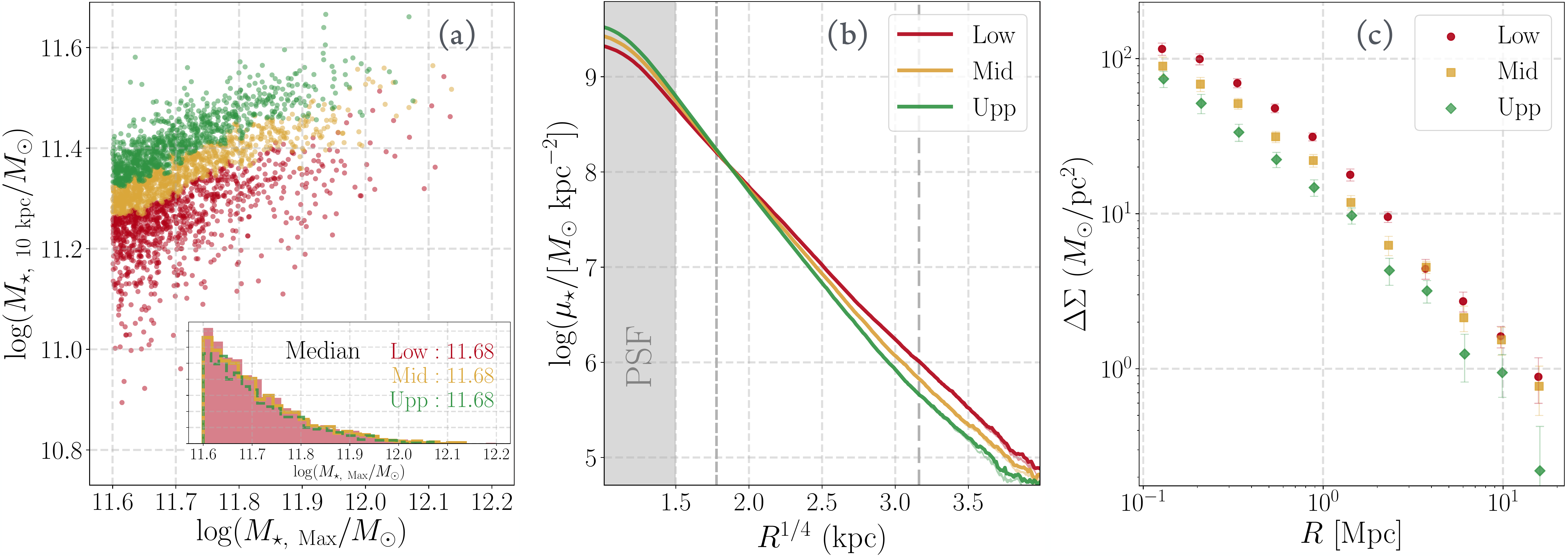}
      \caption{
          (a) Distribution of massive galaxies on the aperture mass plane.  
          We group massive galaxies into three subsamples based on the ranking of their 
          \minn{} at fixed \mmax{}.
          The inset plot demonstrates that the three subsamples share similar 
          distributions of \mmax{}. 
          (b) Median surface mass density profiles of the subsamples visualize the 
          differences in their stellar mass distributions.
          The x-axis employs a $R^{1/4}$ scaling. The shaded region within $\sim 5$ 
          kpc highlights the region affected by seeing.
          Two dashed lines label the 10 kpc (short) and 100 kpc (long) radius.
          (c) The stacked g--g lensing signals prove that at fixed \mmax{}, massive 
          galaxies with lower \minn{} tend to live in more massive dark matter haloes.
          The \texttt{Jupyter} notebook for this figure is available here:
          \href{https://github.com/dr-guangtou/asap/blob/master/note/fig1.ipynb}{\faGithubAlt}
          }
      \label{fig:m100_m10_dsig}
  \end{figure*}

\section{Galaxy--galaxy weak lensing methodology}
    \label{sec:dsigma}      
   
    
    Galaxy--galaxy lensing measures the coherent shape distortion of background 
    galaxies around foreground lens galaxies. 
    Please refer to \citet{HSC-WLCAT} for a detailed description of the construction 
    of our shear catalogue. A detailed description of our method for computing 
    $\Delta\Sigma$ is presented in Speagle et al in prep. 
    Our methodology is briefly summarized below.
    
    The HSC shape catalogue includes a per-galaxy optimal weight defined as
    
    \begin{equation}
        w_i = \frac{1}{e^2_{\rm rms} + \sigma^2_{e,i}}
    \end{equation}
    
    \noindent where $\sigma_{e,i}$ is the shape measurement error per source galaxy 
    and $e_{\rm rms}$ is the intrinsic shape noise.

	We follow the methodology outlined in \citet{Singh2017} to measure the excess 
	surface mass density (hereafter $\Delta \Sigma$) profiles of lens galaxies. 
	Using this method, we measure $\Delta \Sigma$ as:
	
    \begin{equation}
        {\Delta\Sigma}_{\rm LR}(r) = \frac{\Sigma_{\rm Ls} w_{\rm Ls} \gamma_{t}^{(\rm ls)} \Sigma_{\rm crit}^{(\rm Ls)}}{\Sigma_{\rm Ls} w_{\rm Ls}} - \frac{\Sigma_{\rm Rs} w_{\rm Rs} \gamma_{t}^{(\rm Rs)} \Sigma_{\rm crit}^{(\rm Rs)}}{\Sigma_{\rm Rs} w_{\rm Rs}}
    \end{equation}
    
    \noindent where we use $L$ for a real-lens galaxy and $R$ for random point. 
    The superscript or subscript $Ls$ indicates measurement for  
    lens-source pair, while $Rs$ means the measurement for random-source pair. 
    $\gamma$ is the tangential shear, $w$ is the weight, and $\Sigma_{\rm crit}$ is 
    the critical surface density defined as:
    
    \begin{equation}
        \Sigma_{\rm crit}=\frac{c^2}{4\pi G} \frac{D_A(z_s)}{D_A(z_l) D_A(z_l, z_s) (1+z_l)^2}
    \end{equation}
    
    \noindent where $D_A(z_L)$, $D_A(z_s)$, and $D_A{z_L, z_s}$ are the angular 
    diameter distances to lens (random), source, and between them, respectively.  
    We use 11 radial bins uniformly spaced in log-space from 200 kpc to 10 Mpc 
    (physical units are assumed). 
    The redshift distribution of random points is matched to the lens sample.
    
    The subtraction of signal around random positions helps remove overestimated 
    jackknife errors (e.g. \citealt{Clampitt2017, Shirasaki2017}) and
    accounts for non-negligible coherent additive bias of the shear measurements
    (e.g. \citealt{TakadaHu2013}).
    This method has been adopted by the Dark Energy Survey (DES; e.g. 
    \citealt{Prat2017}) and the Kilo-Degree Survey (KiDS; e.g. \citealt{Amon2018}).
    
    
    
    We selected source galaxies based on the following criteria. 
    First, a set of photo--$z$ quality cuts are applied to the sample; these are the
    \texttt{basic} cuts that are described in Speagle\etal~(in prep.). 
    For each lens, we further require $z_{\rm s} - z_{\rm L} \ge 0.1$ and 
    $z_{\rm s} > z_{\rm L} + \sigma_{s,68}$, where $\sigma_{s,68}$ is the 
    1$\sigma$ confidence interval of the source photo--$z$. 
    
    Errors are estimated via jackknife resampling. 
    We divide the S16A \texttt{WLFDFC} footprint into 41 roughly equal-area jackknife 
    regions with regular shapes. 
    In practice, the effective number of jackknife regions varies, depending on the 
    specific subsample of lenses. 
    Typically $N_{\rm JK}>30$. 
    The diagonal errors for $\Delta\Sigma$ are then estimated as:

    \begin{equation}
        \mathrm{Var}_{\rm Jk}(\widehat{\Delta\Sigma}) = \frac{N_{\rm Jk} - 1}{N_{\rm Jk}} \sum\limits_{i=1}^{N_{\rm Jk}} (\Delta\Sigma_{i} - \overline{\Delta\Sigma})^2
    \end{equation}    
    
    \noindent where $N_{\rm Jk}$ is the number of jackknife regions, $\Delta\Sigma_{i}$ 
    is the \dsigma{} profile in each region, and $\overline{\Delta\Sigma}$ is the mean 
    profile among all jackknife regions.

    We measure the stacked \dsigma{} profiles of massive galaxies using a pure Python  
    g--g lensing pipeline designed for the HSC survey: \texttt{dsigma} (available here: 
    \href{https://github.com/dr-guangtou/dsigma}{\faGithub}).    
	Please refer to Speagle\etal{} in prep. for more technical details of \texttt{dsigma}
	and the g--g lensing measurements. 
	           
\section{Measurements}
    \label{sec:measure}
    
\subsection{\mden{} profiles and aperture stellar masses}
    \label{ssec:mprof}      
        
    
    We measure 1-D surface brightness profiles along the major axis of massive galaxies 
    using HSC $i$--band images which typically have the best imaging conditions. 
    We apply an empirical background correction and adaptively mask out neighbouring 
    objects based on their brightness and distance to the target.
    At a given radius, we use the median intensity after 3$\sigma$--clipping 
    along an elliptical isophote twice to measure the surface brightness 
    level\footnote{We use projected 2-D stellar mass maps from hydro-simulation to show
    that our profiles are robust against the impact of unmasked flux from other 
    objects (Ardilla\etal~in prep).~}.
    Using the average \m2l{} measured from SED fitting, we then convert these 
    profiles into surface density profiles of stellar mass -- denoted \mden{}. 
    Integration of the \mden{} profiles provides us with \mstar{} within an 
    elliptical aperture. 
    \citetalias{Huang2018b} contains more technical details about our procedure. 
    
    We can reliably derive \mden{} profiles out to more than 100 kpc for individual 
    massive galaxies without being limited by the background subtraction. 
    On small scales, our profiles are resolved down to 
    $\sim 5$--6 kpc\footnote{1.0\asec{} corresponds to 3.2 and 6.17 kpc at $z=0.19$ 
    and 0.51, respectively; while the mean $i$-band seeing has FWHM $=0$\asec{}$.58$.}.
    
    
    In \citetalias{Huang2018b} and \citetalias{Huang2018c}, we use \mstar{} within 10 
    kpc (\minn{}) and 100 kpc (\mtot{}) as measures of the inner and `total' \mstar{}
    of a galaxy. 
    We also show that \minn{} can be used as a rough proxy for the mass of the 
    \ins{} component. 
    In this work, instead of continuing to use \mtot{}, we choose to use the maximum 
    1-D stellar mass (\mmax{}) as a proxy of `total' \mstar{}. 
    This choice integrates the \mden{} profile to the radius where the median intensity 
    is consistent with the standard deviation of the sky background.
    We have shown that \mmax{} on average adds another 0.03 to 0.05 dex of \mstar{}
    compared with \mtot{}; hence, this approach should bring us a little closer to 
    the true `total' \mstar{}. 
    This choice is motivated by the assumption of the empirical model but does not
    change the key results of this work, which we explain in \S \ref{sec:model}. 
    
    
    As was the case in \citetalias{Huang2018b}, we cannot derive 1-D profiles for 
    $\sim 11$ per cent of massive galaxies due to strong contamination 
    (e.g. a bright star or foreground galaxy) or complex inner structure (e.g. 
    on-going major merger)\footnote{The \mcmodel{} distribution of these galaxies is 
    similar to the whole sample; hence, excluding them should not bias our model.}. 
    Meanwhile, as shown in \citet{SynPipe}, \texttt{hscPipe} tends to 
    classify some stars as extended objects. 
    We find that these contaminations can be easily picked up as outliers on the 
    \mtot{}--\minn{} plane and removed using \logmtot{}$-$\logminn{}$\le 0.03$. 
    In this work we ignore the \m2l{} gradients.  
    Based on \citet{Roediger2015}\footnote{
    $\log(M_{\star}/L_{i}) = 0.83 \times (g-i) - 0.597$ 
    for the \texttt{FSPS} stellar population model}, a color difference of 
    $\Delta (g-i) = 0.2$, which is roughly the average $g-i$ color difference between 
    10 to 100 kpc, translates into a \m2l{} difference of 
    $\Delta \log(M_{\star}/L_{i})\sim 0.15$.
    Considering that the \cmodel{} photometry measures the average color for the main 
    body of massive galaxies, we believe that the systematic uncertainty caused by 
    ignoring the color gradient should smaller than this value. 
    Assuming a negative color gradient, we may be slightly underestimating \minn{} 
    while slightly overestimating \mmax{}. 
   
    
    Our sample contains 38,653 galaxies with \logmmax{}$\ge 11.0$ at 
    $0.19 \le z \le 0.51$.
    Fifty-seven per cent of them have spec--$z$'s.     
    
\renewcommand{\arraystretch}{1.5}
\begin{table*}
    \normalsize 
    \begin{tabular}{c | c | c }
    
    \hline
    \rowcolor{Gray}
    Symbol & Origin & Explanation \\
    \hline
    
    \rowcolor{CadetBlue}
    \minn{}   &   Observation  &   Aperture \mstar{} within inner 10 kpc via integrating the 1-D \mden{} profile \\
    \rowcolor{CadetBlue}
    \mmax{}   &   Observation  &   Maximum aperture \mstar{} via integrating the 1-D \mden{} profile \\ \hline
    
    \rowcolor{Dandelion}
    \mvir{}   &  \smdpl{}     &   Dark matter halo mass within virial radius (\citealt{BryanNorman1998}) \\
    \rowcolor{Dandelion}
    \mpeak{}  &  \smdpl{}     &   Peak historical dark matter halo mass \\ \hline
       
    \rowcolor{LimeGreen}
    \mall{}   &  \um{}        &   \mstar{} of all galaxies (central and satellites) within a host dark matter halo \\ 
    \rowcolor{LimeGreen}
    \mgal{}   &  \um{}        &   \mstar{} of a central or satellite galaxy \\     
    \rowcolor{LimeGreen}
    \mcen{}   &  \um{}        &   \mstar{} of the central galaxy \\ 
    \rowcolor{LimeGreen}
    \mins{}   &  \um{}        &   \mstar{} of the \insitu{} component of a galaxy \\ 
    \rowcolor{LimeGreen}
    \mexs{}   &  \um{}        &   \mstar{} of the \exsitu{} component of a galaxy \\ 
    \rowcolor{LimeGreen}
    $\delta_{\rm gal}$   &  \um{}  &  \mgal{}$/$\mall{}: stellar mass fraction of galaxy in the halo  \\         
    \rowcolor{LimeGreen}
    $\delta_{\rm ins}$   &  \um{}  &  \mins{}$/$\mgal{}: stellar mass fraction \ins{} component in the galaxy  \\      
    \rowcolor{LimeGreen}
    $\delta_{\rm exs}$   &  \um{}  &  \mexs{}$/$\mgal{}: stellar mass fraction \exs{} component in the galaxy  \\  \hline 

    \rowcolor{Melon}
    \mallmod{} &  \asap{}  & Model predicted total \mstar{} in a host dark matter halo  \\ 
    \rowcolor{Melon}
    \mgalmod{} &  \asap{}  & Model predicted \mstar{} of a galaxy (central or satellite) \\ 
    \rowcolor{Melon}
    \mcenmod{} &  \asap{}  & Model predicted \mstar{} of a central galaxy \\ 
    \rowcolor{Melon}
    \minsmod{} &  \asap{}  & Model predicted \insitu{} \mstar{} \\     
    \rowcolor{Melon}
    \mexsmod{} &  \asap{}  & Model predicted \exsitu{} \mstar{}  \\ 
    \rowcolor{Melon}
    \minsten{} &  \asap{}  & Model predicted \insitu{} \mstar{} within inner 10 kpc \\      
    \rowcolor{Melon}
    \mexsten{} &  \asap{}  & Model predicted \exsitu{} \mstar{} within inner 10 kpc \\  
    \rowcolor{Melon}
    \minnmod{} &  \asap{}  & Model predicted 10 kpc aperture \mstar{} \\ 
    \hline
    
    \end{tabular}
    
    \caption{
        Definitions of halo masses and stellar masses used in this work. 
        Rows with different colors are used to separate the masses defined 
        in observations (blue), in (\smdpl{}) dark matter simulation (orange), 
        in the \um{} model predictions (green), and in the Accelerated SAP (\asap{}) model 
        developed in this work (red). 
        Notation with calligraphic letters is used to denote model predictions 
    }
    \label{tab:mass}
    
\end{table*}

  \begin{figure*}
      \centering 
      \includegraphics[width=16cm]{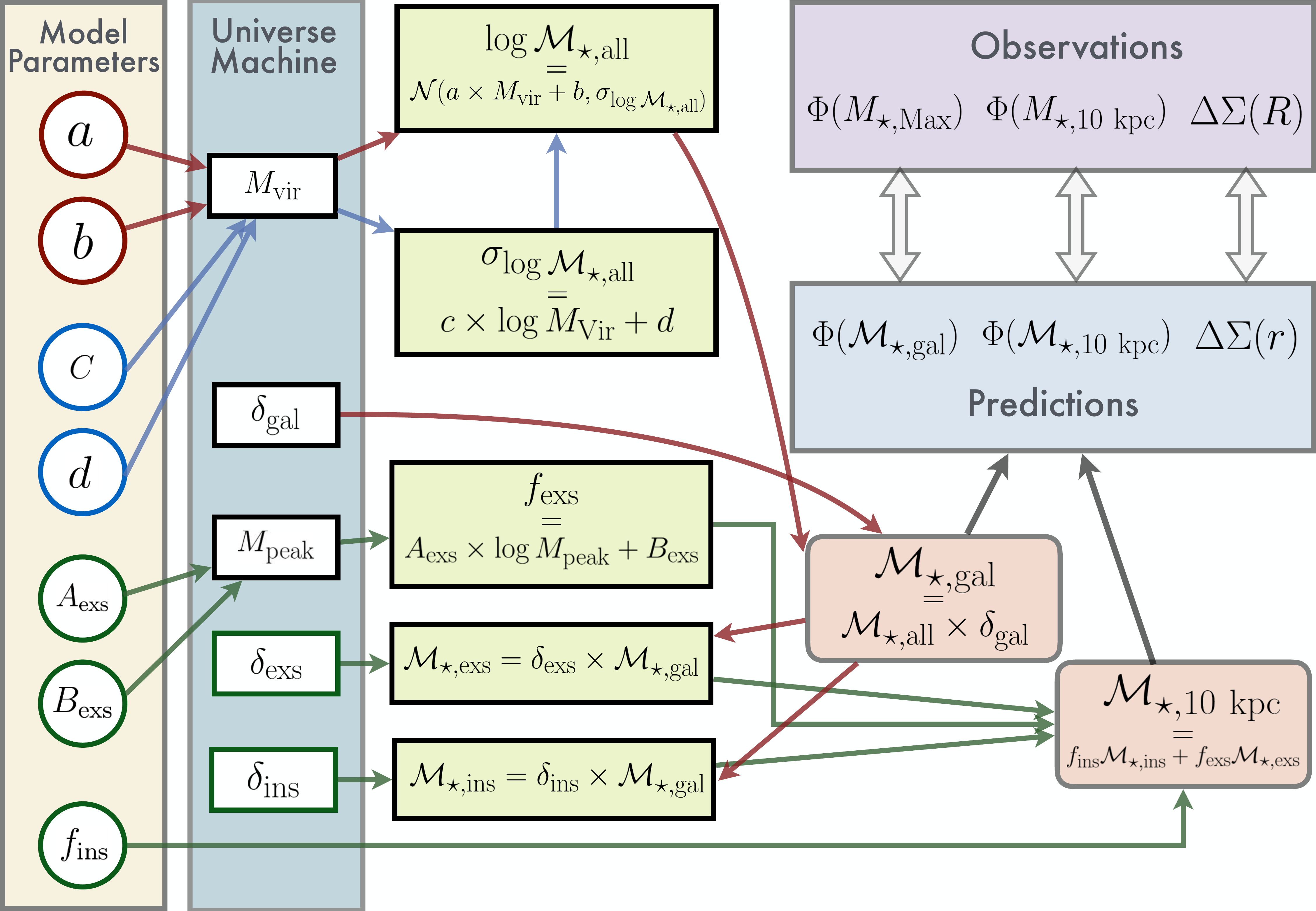}
      \caption{
          Flowchart for the basic design of the \asap{} model.  
          The \um{} predictions adopted in this model are highlighted on the right. 
          These correspond to: \mpeak{}--the peak halo mass; \fgal{}--the ratio between 
          the stellar mass of a galaxy (\mgal{}) and the total stellar mass within 
          the halo (\mall{}); also the fraction of \ins{} (\fins{}) and \exs{} 
          (\fexs{}) components in the \mstar{} of each galaxy.
          The seven free model parameters are labelled on the bottom: $a$ and $b$
          describe a $\log$--$\log$ linear relation between \mpeak{} and \mall{}; 
          $c$ and $d$ describe a linear relation between the scatter of \mall{}
          and \mpeak{}.  
          These four parameters, along with the \fgal{} fraction predicted by \um{}, 
          provide predictions of the stellar mass of each galaxy (\mgalmod{}) that will 
          be compared with the observed \mmax{}. 
          The stellar mass of \ins{} and \exs{} components (\mexsmod{} and \mexsmod{}) 
          are estimated using \mgalmod{}, \fins{}, and \fexs{}.
          The predicted stellar mass in 10 kpc (\minnmod{}) requires another three
          free parameters; 
          $f_{\rm ins}$ describes the fraction of \ins{} stars located within the 
          inner 10 kpc, and the fraction of \exs{} stars in 10 kpc follows a linear 
          relation with \logmpeak{} that is characterized by $A_{\rm ins}$ and
          $B_{\rm exs}$.
          A keynote version of this flowchart is available here:
          \href{https://github.com/dr-guangtou/asap/blob/master/doc/flowchart.key}{\faGithubAlt}
          }
      \label{fig:asap_flowchart} 
  \end{figure*}
    
\subsection{Stellar mass functions}
    \label{ssec:smf}
    
     
    In this work, we estimate the stellar mass function (SMF) of \mmax{} 
    ($\Phi_{\rm max}$) in seven bins between $11.6 \le$\logmmax{}$<12.3$, 
    while we estimate the SMF of \minn{} ($\Phi_{\rm 10}$) in ten bins between 
    $10.8 \le$\logminn{}$<11.8$.
    We separate the current \texttt{WLFDFC} area into 30 smaller regions, and derive 
    uncertainties via jackknife resampling. 
    We add a 10 per cent uncertainty to represent the potential impact of galaxies 
    without a useful 1-D profile. 
    We take the uncertainty of \mstar{} measurements into account by integrating 
    the normalized posterior distribution function (PDF) of the \mstar{} of each 
    galaxy\footnote{We assume that the PDF is described by a Gaussian distribution.} 
    to estimate its contribution in each \mstar{} bin. 
    For a given \mstar{} bin with lower and upper boundary of $M_l$ and $M_u$, the 
    effective number of galaxies in the bin is: 

    \begin{equation}
        N_{\rm eff} = \sum_{i=1}^{n_{\rm gal}} \frac{1}{2}[\mathrm{erf}(\frac{M_u - M_i}{\sqrt{2}\sigma_i}) - \mathrm{erf}(\frac{M_l - M_i}{\sqrt{2}\sigma_i})]
    \end{equation}    
    
    \noindent where $M_i$ is the mean \mstar{} and $\sigma_i$ is the uncertainty for 
    each massive galaxy and $\mathrm{erf()}$ is the error function.
        
    
    By comparing our results with SMFs from the PRIsm MUlti-object Survey 
    (\href{https://primus.ucsd.edu}{PRIMUS}; e.g. \citealt{Moustakas2013}) at a 
    similar redshift, we find that massive galaxies with \logmmax{}$\ge 11.6$ are a mass 
    complete sample and are considered in the following modelling. 
    In total, we have \textbf{6481 and 3156 galaxies} at \logmmax{}$\ge 11.5$ and 
    $\ge 11.6$; \textbf{5756 and 2944} of them have spec--$z$. 
    The \mmax{}--\minn{} distribution of our sample is shown in 
    Figure~\ref{fig:m100_m10_dsig} and Figure~\ref{fig:asap_model_1}.
    The SMFs of \mmax{} and \minn{} for the \logmmax{}$\ge 11.6$ sample are shown
    in panel (b) of Figure~\ref{fig:asap_model_1}.
    
    
    The SMFs of \mmax{} and \minn{} are highly correlated as \minn{} is included in the 
    measurement of \mmax{}.
    We calculate the covariance matrix of the joint \mmax{}--\minn{} SMF using the same 
    jackknife samples. 
    
\subsection{Galaxy--galaxy lensing signals across the aperture mass plane}
    \label{sec:plane}
    
    
    In \citetalias{Huang2018c} (see Figure~3), we find that massive central galaxies 
    of \redm{} clusters (e.g. \citealt{Rykoff2014, Rozo2014}) have lower 
    \minn{} than those in less massive haloes at fixed \mtot{}, which suggests that
    the stellar mass distributions in massive central galaxies depend on their 
    \mhalo{}. 
    Therefore, we expect a gradient of \mhalo{} across the aperture mass plane. 
    Our goal is to map out this gradient directly using weak lensing and without 
    relying on any \redm{} cluster catalog.
        
    
    Panel (a) in Figure~\ref{fig:m100_m10_dsig} shows the distribution of massive
    galaxies over the \mmax{}--\minn{} plane. 
    We group galaxies into three sub-samples based on the ranking of their \minn{} at 
    fixed \mmax{}, following a similar strategy employed in \citet{Mao2018}. 
    As illustrated in the inset panel, the three sub-samples share almost identical 
    distributions of \mmax{}.
    Therefore, they represent massive galaxies with different stellar mass 
    distributions at the same `total' stellar mass, as proved by their median \mden{} 
    profiles (panel (b) of Figure~\ref{fig:m100_m10_dsig}). 
    Galaxies with lower \minn{} have lower \mden{} on small radial scales and have 
    larger extended outer envelopes. 
    The median \mden{} profiles cross each other at $\sim 12$-15 kpc, close to the 
    effective radius ($R_{\rm e}$) of these galaxies.
    
    
    We then measure the stacked \dsigma{} profiles of these three sub-samples using 
    the method described in \S\ref{sec:dsigma}. 
    The results are displayed in panel (c) of Figure~\ref{fig:m100_m10_dsig}. 
    It is very clear from this Figure that, on average, massive galaxies with lower 
    \minn{} have higher \dsigma{} signals indicating that they live in more massive 
    dark matter haloes. 
    This confirms the expected trend across the aperture mass plane that was first 
    identified in \citetalias{Huang2018c} using broad \mhalo{} bins from cluster
    catalog. 
   
    
    Thanks to the impressive weak lensing capabilities of the HSC survey, we can 
    further group massive galaxies into bins of \mmax{} and \minn{} and investigate 
    the variation of their stacked \dsigma{} profiles and halo masses. 
    The \texttt{Jupyter} notebook for measuring these \dsigma{} profiles can be
    found here: 
    \href{https://github.com/dr-guangtou/asap/blob/master/note/hsc_weak_lensing.ipynb}{\faGithubAlt}.
    We also make a \texttt{GIF} animation to visualize this variation:
    \href{https://github.com/dr-guangtou/asap/blob/master/doc/dsig_over_aperture_plane.gif}{\faImage}
    
    To account for scatter in \mhalo{} within each \mmax{}--\minn{} `box', the impact 
    of satellites, and the two-halo term, we model our lensing signals using a full 
    forward model based on N-body simulations and a state-of-the-art semi-empirical 
    model. 
    We will group our massive galaxies into 12 bins of aperture masses while making
    sure that (1) there are enough massive galaxies in each bin so that the stacked 
    \dsigma{} profile has good \s2n{}; and (2) the \minn{} bins at fixed \mmax{}
    represent massive galaxies with different stellar mass in the inner region.
    We explain the details of the model in Section \ref{sec:model}.
    
  \begin{figure*}
      \centering 
      \includegraphics[width=16cm]{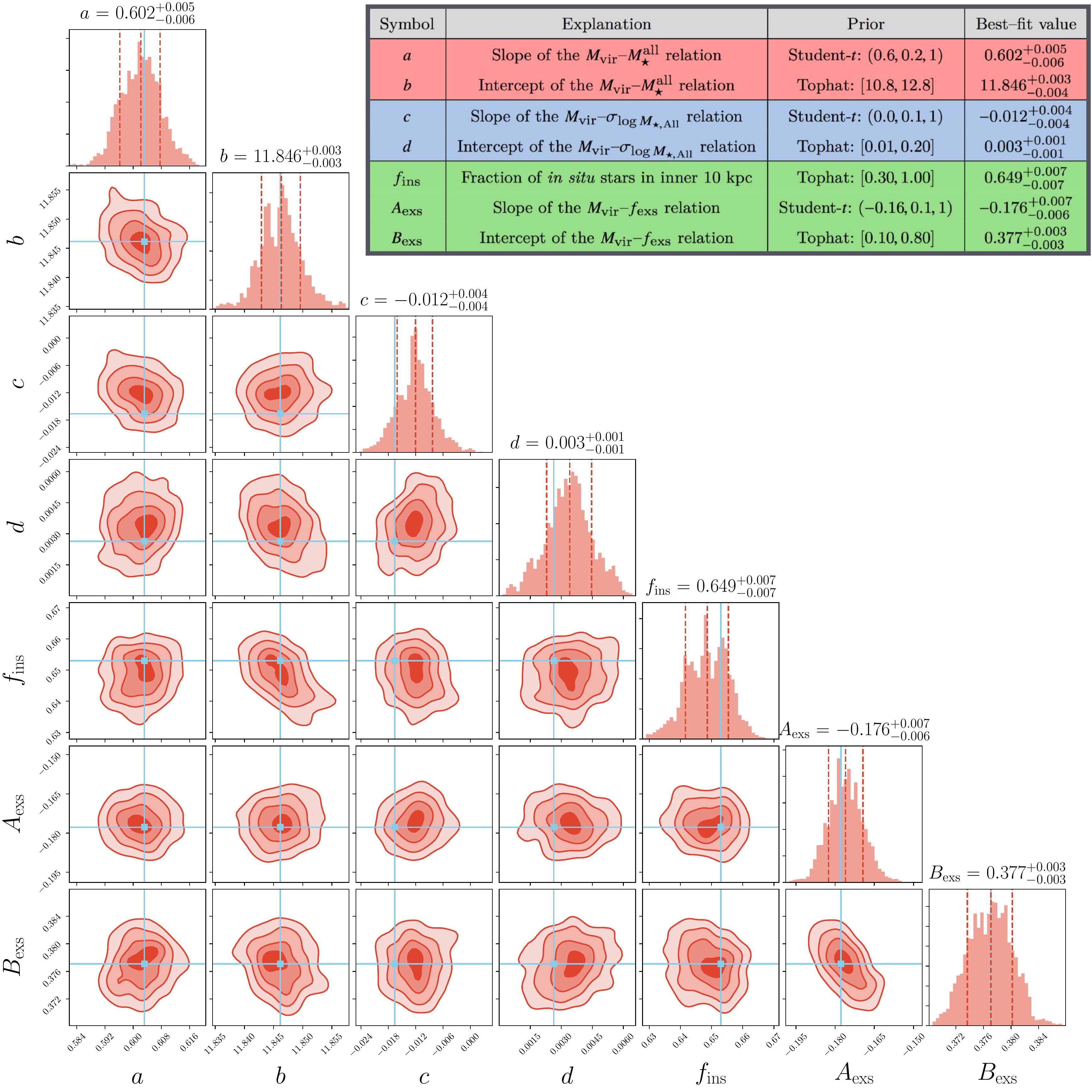}
      \caption{
          Corner plot for the posterior probability distributions of parameters in 
          the model. 
          The contour levels describe 68 per cent, 95 per cent, and 99.7 per cent 
          enclosed probability regions.
          The explanations, ranges of uniform priors, and the best-fit values along
          with their uncertainties are highlighted in the upper-right table.
          The colors separate parameters into three groups as indicated in the 
          flowchart.
          The \texttt{Jupyter} notebook for this figure is available here:
          \href{https://github.com/dr-guangtou/asap/blob/master/note/fig3_default.ipynb}{\faGithubAlt}
          }
      \label{fig:asap_corner}
  \end{figure*}

  \begin{figure*}
      \centering 
      \includegraphics[width=\textwidth]{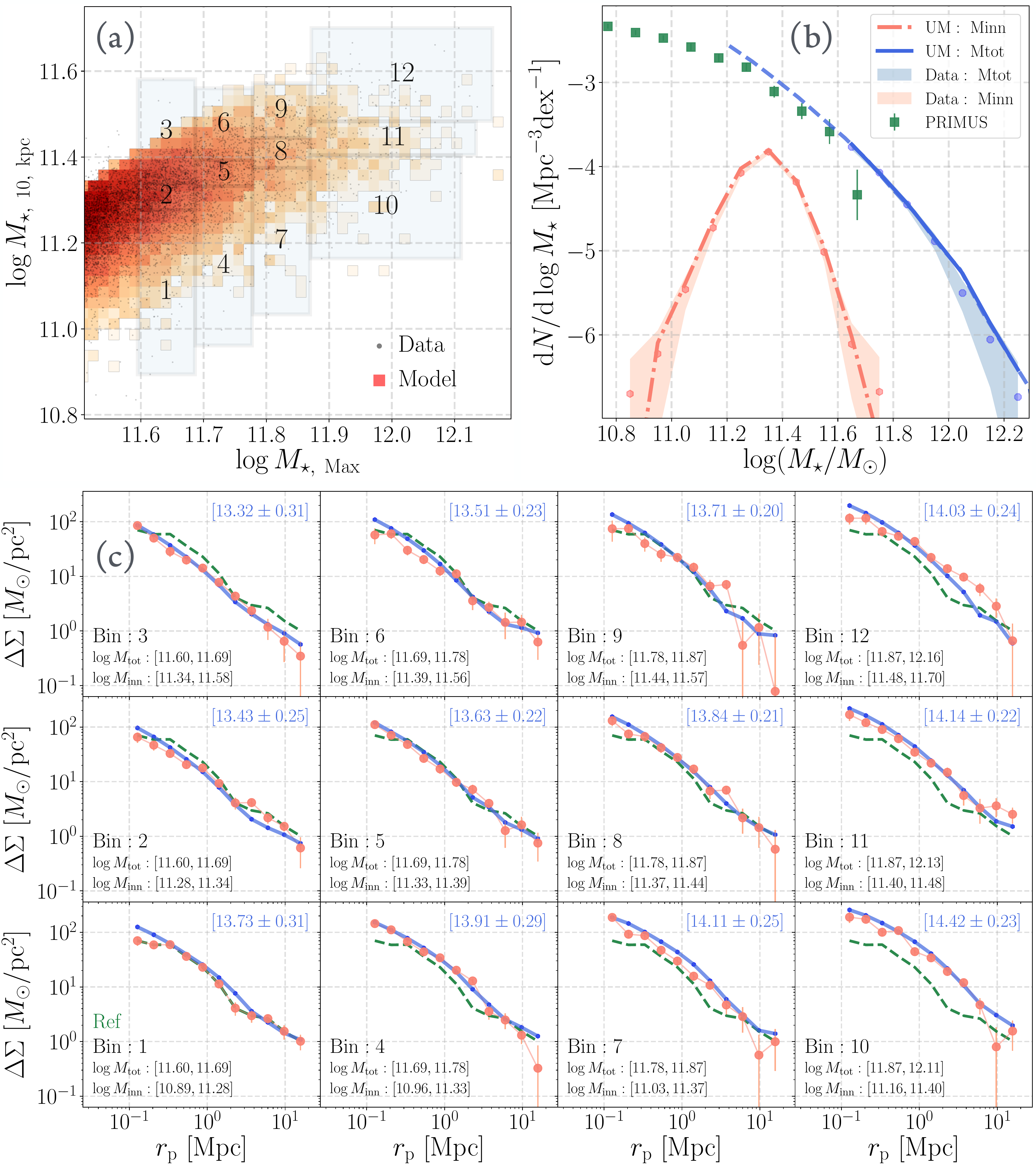}
      \caption{Performance of the best-fit model.
          (a) Comparison between the observed HSC galaxies (grey points) and modeled 
          galaxies (density plot) over the \mmax{}--\minn{} plane.
          Both observed and modeled galaxies with \logmmax{}$\ge 11.6$ are grouped 
          into the same 12 bins using \mmax{} and \minn{} values to compare the g--g 
          lensing signals within each bin.
          (b) Comparisons of observed (dots and shaded regions) and modeled SMFs 
          (solid lines) for \mmax{} (blue) and \minn{} (red). 
          We also overplot the SMF from the PRIMUS survey at similar redshift to 
          show the shape of SMF at lower \mstar{}.
          (c) Comparisons of g--g lensing signals in each \mmax{}--\minn{} bin. 
          The bin number and the mass range of \mmax{} and \minn{} of each bin is 
          given in the lower-left corner of each subplot.  
          The observed g--g lensing signals are shown as red points, while the blue 
          lines show the modeled lensing signal. 
          The weak lensing signal from $\mathrm{Bin=1}$ (bottom-left plot) is shown 
          in each subplot as a green dashed line to highlight the evolution of 
          $\Delta\Sigma$ amplitudes across various bins. 
          The median \mvir{} in each bin is shown in
          the upper-right corner of each subplot.
          The \texttt{Jupyter} notebook for this figure is available here:
          \href{https://github.com/dr-guangtou/asap/blob/master/note/fig4.ipynb}{\faGithubAlt}
          }
      \label{fig:asap_model_1}
  \end{figure*}

\section{Modelling the \mhalo{}-\mmax{}-\minn{} relation}
    \label{sec:model}

\subsection{Goals of the Model}

    Our goal is to construct a model that connects the hierarchical growth of dark 
    matter halos to the assembly and structure of high mass central galaxies. 
    
    
    Ideally, we could directly compare with predictions from hydro-dynamical 
    simulation, such as \href{http://www.illustris-project.org}{Illustris} 
    (e.g. \citealt{Vogelsberger2014, Genel2014}) 
    or \href{http://icc.dur.ac.uk/Eagle/}{EAGLE} (e.g. \citealt{Schaye2015, Crain2015}) 
    that are being used to study the evolution of massive galaxies 
    (e.g. \citealt{Wellons2016b, RodriguezGomez2016, Qu2017}). 
    However, current hydro-simulations typically lack of the volume to study galaxies 
    at high--\mhalo{} end statistically. 
    In addition, we also want a model with the flexibility (free parameters) to 
    fit the actual observations. 
    
    An alternative approach would be to use a semi-analytic model (SAM) based on 
    dark matter simulations and approximate physical recipes 
    (e.g. \citealt{WhiteFrenk1991, Benson2010, Guo2011, 
    Henriques2015, Somerville2015, Croton2016}) could be another approach. 
    However, while recent progress has been made in this area, fitting the large 
    numbers of parameters that a SAM typically uses is still non trivial
    (e.g. \citealt{Lu2011, Benson2014, Benson2017}).
    
    For these reasons, we base our formalism on the recently developed semi-empirical 
    model approach (e.g. \citealt{Becker2015, RodriguezPuebla2017, Moster2018, 
    Behroozi2018}). 
    This new methodology makes rather minimal \emph{a priori} assumptions about the 
    galaxy--halo connection, and is constrained by observations at different redshifts 
    (stellar mass growth, star-formation history, and clustering properties of galaxies 
    across a wide range of halo masses and redshifts). 
    This results in a model that can predict the properties of 
    \emph{individual galaxies} and how they connect with the full assembly history of 
    their dark matter halos.

\subsection{Simulations and \texttt{UniverseMachine} framework}
    \label{sec:um_smdpl}
 
    
    \um{} (\citealt{Behroozi2018}; code available here: 
    \href{https://bitbucket.org/pbehroozi/universemachine}{\faBitbucket})
    is a massively parallel implementation of a semi-empirical modeling method.
    It is capable of reproducing key observations (e.g. stellar mass functions, 
    star formation rates, and quenched fractions) over a large range of stellar masses
    and redshifts.  
    For a given halo from a cosmological simulation, \um{} parametrizes its star 
    formation rate (SFR) as a function of halo mass, halo accretion rate, and redshift. 
    \um{} exploits the Markov Chain Monte Carlo (MCMC) Bayesian method to compare 
    results with a series of compiled observations.   
    
    
    The \um{} model we use here is based on the 
    \href{https://www.cosmosim.org/cms/simulations/smdpl/}{Small MultiDark Planck} (\smdpl{})
    simulation, 
    which is part of the MultiDark simulation series using a Planck cosmology. 
    It has a 400 Mpc$/h$ simulation box size and uses $3840^3$ particles.
    The dark matter mass resolution is $10^8 M_{\odot}/h$.
    The volume of the \smdpl{} simulation is two times larger than the volume from 
    which our HSC sample at $0.19 \le z \le 0.51$ is drawn from. 
    Dark matter subhalo properties are extracted using the 
    \href{https://bitbucket.org/gfcstanford/rockstar}{\rockstar{}} (\citealt{Rockstar}) halo 
    finder with merger trees generated by the 
    \href{https://bitbucket.org/pbehroozi/consistent_trees}{\texttt{Consistent Trees}} code. 
    Halo mass is defined as the mass within the virial radius (\mhalo{}) using 
    the formula from \citet{BryanNorman1998}. 
    For satellite galaxies, we will also use their peak \mhalo{} over their accretion 
    history (\mpeak{}). 
    Here we use the snapshot at $z\sim 0.37$, which is very close to the mean redshift 
    of our sample ($\overline{z}\sim 0.32$).
                
    
    The fiducial \um{} model predicts a `galaxy mass' and an `ICL' mass. 
    During mergers, a fraction of stars from the incoming satellite become unbound 
    by the gravitational well of the galaxy and are added to the `ICL' component.
    Although there is evidence for an unbound diffuse stellar component around 
    nearby massive galaxies (e.g. \citealt{Kelson2002, Bender2015, Longobardi2015}), 
    the main motivation of this approach is to make sure the SMF matches 
    observational constraints at low redshift, otherwise \um{} over-produce
    the SMF at the high-\mstar{} end (Behroozi\etal{}2018). 
    However, as we showed in \citetalias{Huang2018b}, it is extremely difficultly to 
    photometrically separate out a physically meaningful `ICL' component. 
    More importantly, the ICL component is also an integrated part of the assembly 
    history of massive galaxies and should be taken into account when studying 
    their galaxy--halo connection. 
    
    Therefore, instead of using the `galaxy' and `ICL' separation, we use a 
    specially tailored \um{} model that provides a more physically 
    motivated decomposition of stars in massive galaxies: for each galaxy, our 
    \um{} model will predict the mass of the \ins{} and \exs{} components
    (\mins{} and \mexs{}). 
    As mentioned earlier, these are stars formed inside and outside the \emph{main 
    progenitor} of the subhalo. 
    For each galaxy, the stellar mass of the galaxy (\mgal{}) is simply the sum
    of \mins{} and \mexs{}.  
    The stellar mass of the central galaxy in a halo is denoted as \mcen{}. 
    For each halo, we also calculate the total stellar mass within the halo (\mall{}) 
    meaning the sum of stellar mass of the central and \emph{all} satellites.
    These stellar mass definitions are given in Table~\ref{tab:mass}.

\subsection{\asap{} model}
    \label{sec:um_asap}
     
    
    In this section, we explain the design and key assumptions behind our empirical 
    model, which we call the \asap{}\footnote{An initialism for Alexie Leauthaud, 
    Song Huang, Andrew Hearin, and Peter Behroozi, the first names of the main 
    contributors.} model.  
    Constrained by the observed SMFs of different aperture masses and \dsigma{} 
    profiles across the aperture mass plane, the \asap{} model will connect
    \mhalo{}, \mins{}, and \mexs{} to the observed stellar mass distributions among 
    massive galaxies. The \asap{} model is based on the following two key ingredients:
            
    \begin{enumerate}
    
        \item There is a tight $\log-\log$ linear relation between halo mass and the 
              total stellar mass within the halo (TSHMR) at the high-\mhalo{} end 
              (Bradshaw et al in prep). 
                      
        \item The \um{} model provides \ins{} and \exs{} components -- we add a 
              prescription to describe the spatial distributions of these components. 
    
    \end{enumerate}
    

\subsubsection{Total Stellar--Halo Mass Relation (TSHMR)}
    \label{sec:tshmr}
    
    The SHMR is the relation between halo mass and central galaxy mass. 
    Whereas the SHMR has an exponentially rising slope and large scatter at the high 
    mass end, recent hydro-simulations (e.g. \citealt{Pillepich2017a}) and 
    semi-empirical models (e.g. \citealt{Behroozi2018}, Bradshaw\etal{} in prep.) 
    suggest that the TSHMR follows the simple tight, log-linear correlation with 
    \mhalo{} (at least at the high--\mhalo{} end). 
    In observations, the total K$s$--band luminosity or stellar mass in galaxy groups 
    and clusters also show tight relation with halo mass 
    (e.g. \citealt{Lin2004, Ziparo2016, Leauthaud2012b, vanderBurg2014, Budzynski2014, 
    Patel2015, Kravtsov2018}). 
    Motivated by this, we place the TSHMR at the core of our approach.  
    The SHMR then emerges as a consequence of the TSHMR and the assembly histories of 
    halos (e.g. Bradshaw\etal{} in prep.).
    
    We assume a log-linear relation between the mass of the host dark matter halo 
    (\mvir{}) and the total stellar mass within the halo (including the central 
    galaxy, satellites from all subhaloes, and the ICL component; \mallmod{}). 
    The TSHMR used in the \asap{} model is described as: 
    
    \begin{equation}
        \log \mathcal{M}_{\star}^{\mathrm{all}} = a \times (\log M_{\rm vir} - 13.5) + b 
    \end{equation}   
    
    \noindent{} The slope ($a$) and intercept ($b$) are free parameters in our model. 
    We adopt a pivot \mhalo{} of \logmh{}$=13.5$ in all 
    $\log$--linear scaling relations involving halo mass to reduce the degeneracy 
    between the slope and intercept. 
    The exact value of this pivot mass does not impact our results.
    The scatter in this relation is also modeled as a simple linear relation:

    \begin{equation}
        \sigma_{\log\mathcal{M}_{\star}^{\mathrm{all}}} = c \times (\log M_{\rm vir} - 13.5) + d 
    \end{equation}    
     
    \noindent{}where $c$ and $d$ are two additional parameters. 
    The above relations determine the total amount of \mstar{} in each `parent' 
    halo in the \asap{} model. 
    We do not use the value of \mallmod{} directly from the \um{} because the 
    current version of the \um{} is constrained to match stellar mass functions 
    from \citet{Muzzin2013} that contain larger numbers of massive galaxies 
    (see \citealt{Behroozi2018}). 
    We should point out that, when comparing to observations, the scatter should 
    be a combination of the intrinsic scatter of the TSHMR and the measurement 
    errors of observed stellar mass and weak lensing profiles. 
    We will discuss the scatter of TSHMR further in \S \ref{ssec:tshmr}. 
   
    So far, our model has simply `pasted' \mallmod{} values on halos in our 
    simulations. 
    The information that we adopt from the \um{} is the following. 
    The \um{} model tells us, for a given \mallmod{}, how mass is divided up 
    among galaxies. 
    For every galaxy, we compute \fgal$\equiv$\mgalmod{}$/$\mallmod{}. 
    At this stage, we also forward model  uncertainties associated 
    with stellar mass measurements 
    Thus, each galaxy in our mock is assigned a mass following:
        
    \begin{equation}
        \log \mathcal{M}_{\star, {\rm gal}} \sim \mathcal{N}(\log (\mathcal{M}_{\star}^{{\rm all}} \times \delta_{\rm gal}),\ \sigma_{\log \mathcal{M}_{\star}^{\rm all}})
    \end{equation}
   
    \noindent where $\mathcal{N}(\mu, \sigma)$ is a normal distribution with 
    mean value of $\mu$ and standard deviation of $\sigma$.

    We apply this model to both centrals and satellite galaxies. 
    Massive satellites are included in our forward modeling process because we do not 
    attempt to distinguish centrals and satellites in our HSC sample\footnote{
    Uncertainties of photometric redshifts make it difficult to accurately separate 
    centrals and satellites.  
    Meanwhile, the satellite fraction at \logmmax{}$\ge 11.5$ is 
    less than $<10$ per cent; see Sallaberry\etal{} in prep.}. 
   

\subsubsection{Spatial distributions of \ins{} and \exs{} stars}
    \label{sec:mass_prof}   
    
    For every galaxy, the second ingredient that we inherit from the \um{} is the 
    fraction of \ins{} and \exs{} component (\fins{} and \fexs{}). 
    We now model the observed aperture masses, \minn{} and \mmax{}, via a prescription 
    that describes the spatial distributions of \ins{} and \exs{} stars.
    First, we assume the observed \mmax{} is a good proxy for the `total' stellar mass 
    of the galaxy:
    
    \begin{equation}
        \mathcal{M}_{\star}^{\rm max}=\mathcal{M}_{\star}^{\rm ins} + \mathcal{M}_{\star}^{\rm exs}.
    \end{equation} 
    
             
    
    
    Next, we predict \minn{} using two assumptions. 
    First, we assume that a fixed fraction of the \ins{} component is within the 
    inner 10 kpc of the galaxy:
    
    \begin{equation} 
        \mathcal{M}_{\star}^{\mathrm{in,\ 10}} = f_{\rm ins} \times \mathcal{M}_{\star}^{\mathrm{ins}}.
    \end{equation}
     
    Second, we assume that the fraction of \exs{} stars within 10 kpc depends on halo 
    mass: 
    
    \begin{equation}
        \mathcal{M}_{\star}^{\mathrm{ex,\ 10}} = f_{\rm exs} \times \mathcal{M}_{\star}^{\mathrm{exs}}
    \end{equation}
    
    where the relation between the fraction and halo mass is described by: 
   
    \begin{equation}
        f_{\rm exs} = A_{\rm exs} \times (M_{\rm vir} - 13.5) + B_{\rm exs}.
    \end{equation}
       
    Given these two assumptions, the predicted 10 kpc aperture mass is then: 
    \minnmod{}$=$\minsten{}$+$\mexsten{}. 
    To model \minn{} therefore requires three extra free parameters: 
    $f_{\rm ins}$, $A_{\rm exs}$, and $B_{\rm exs}$. 
 
    For satellite galaxies, we use \mpeak{} instead of $M_{\rm vir}$. 
    However, because the fraction of satellite galaxies that are massive enough to be
    included in our sample is very low at the high stellar mass end 
    (Sallaberry et al in prep), this choice has no impact on our results.
    As the scatter of the TSHMR is designed to carry both intrinsic scatter and 
    measurement uncertainties of stellar mass, the predicted \minnmod{} and 
    \mmaxmod{} will be described by normal distributions with the same scatter.

    In total, our model has seven free parameters: two for the TSHMR; two for the 
    scatter of the TSHMR; and three for the fraction of \ins{} and \exs{} stars 
    within 10 kpc. 
    Figure \ref{fig:asap_flowchart} is a visualization of our model.
    
\subsubsection{Predictions for the SMFs and \dsigma{} profiles}
    \label{sec:predict}    
    
    We predict the SMFs of \minnmod{} and \mmaxmod{} using the same method and in the 
    same stellar mass bins for the observed SMFs. 
	Uncertainty in stellar mass measurements is accounted for according to equation (7). 
    
    When comparing the predicted and observed SMFs, we jointly constrain the 
    $\Phi_{\rm max}$ and $\Phi_{\rm 10}$ (referred to as $\Phi_{\rm obs}$) by taking 
    the measured covariance matrix ($\textbf{C}_{\rm obs}$) into account. 
    The log-likelihood for SMF is: 

    \begin{equation}
        \ln \mathcal{L}_{\rm SMF} = -\frac{1}{2} [\Phi_{\rm mod} - \Phi_{\rm obs}]^{T} \textbf{C}_{\rm obs}^{-1} [\Phi_{\rm mod} - \Phi_{\rm obs}] + K   
    \end{equation}
    
    \noindent{}where $\Phi_{\rm mod}$ is the predicted SMFs for \minnmod{} and 
    \mmaxmod{} aligned in the same order with the observed SMFs.
    $K$ is a constant described by 
    $-\frac{1}{2}[\ln(2\pi)N + \ln(\mathrm{det}(\textbf{C}_{\rm obs}))]$
    and $N=17$, which is the total number of mass bins. 

    The lensing observable, \dsigma{}, is computed directly from the simulation using 
    50 million randomly selected dark matter particles and the 
    \texttt{mock\_observables.delta\_sigma} function in the 
    \href{https://github.com/astropy/halotools}{\htools{}} (\citealt{Hearin2017}). 
    We predict the weighted-mean \dsigma{} profiles in the same 12 aperture mass 
    bins used for observation for comparison after considering the uncertainties
    of the predicted \minnmod{} and \mmaxmod{} into the weight. 
    Our method accounts for the effects of scatter, the finite width of our bins, 
    satellite galaxies, and the two-halo term. 
    We ignore the contribution of \mstar{} to \dsigma{} because it is negligible on 
    the scales that we consider ($r > 200$ kpc).

    The log-likelihood for comparing \dsigma{} profiles is described as:
    
    \begin{equation}
        \ln \mathcal{L}_{{\Delta\Sigma}_{j}}= -\frac{1}{2} \sum_{i}^{n} \frac{(\Delta\Sigma_{\rm mod,i}-\Delta\Sigma_{\rm obs,i})^2}{\sigma_{i}^2} + \sum_{i}^{n} \ln(2\pi\sigma_{i}^2)
    \end{equation}
    
    \noindent{}where the sum over $i$ is for $n=11$ radius bins of each \dsigma{} 
    profile and $\sigma_{i}$ is the associated observational uncertainty derived 
    using a jackknife resampling method. 
        
           
        

  \begin{figure*}
      \centering 
      \includegraphics[width=\textwidth]{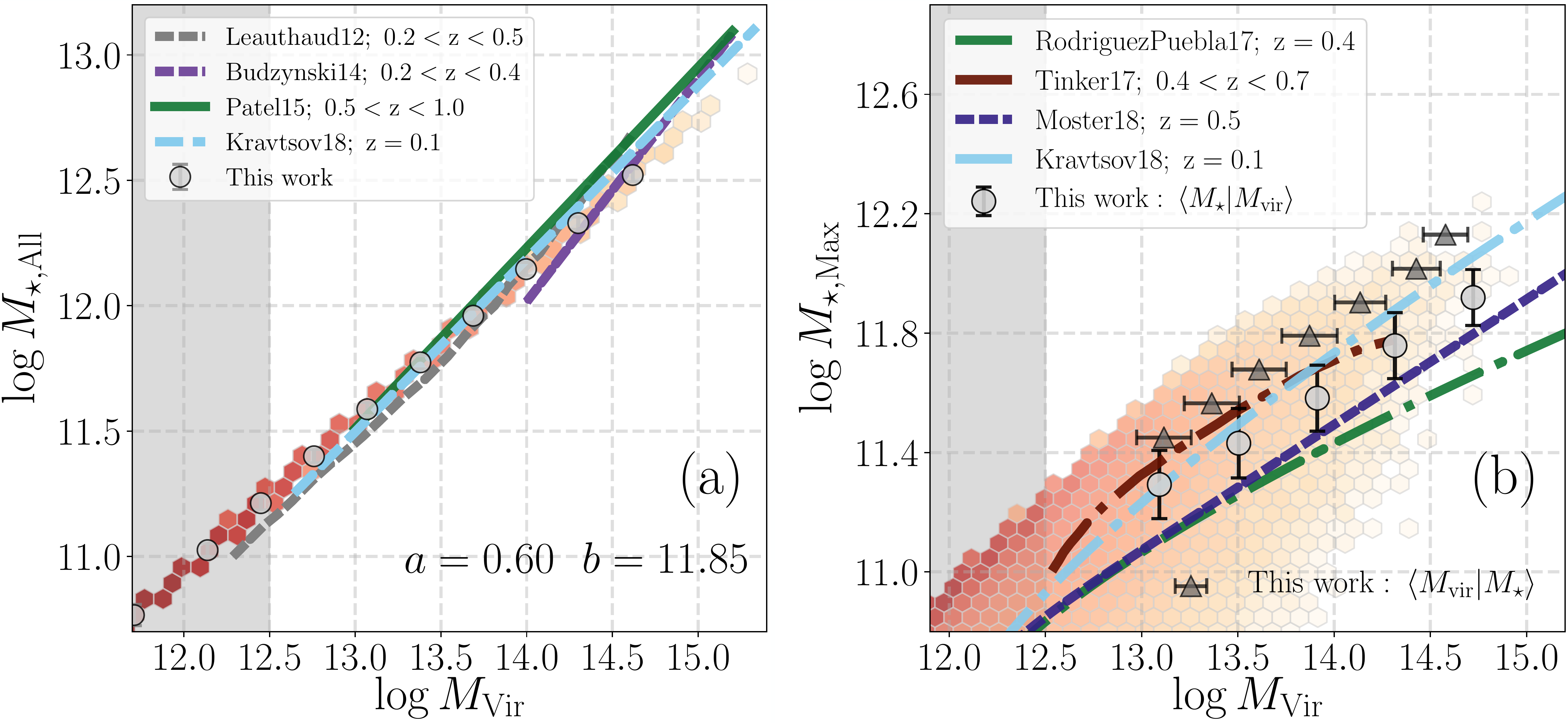}
      \caption{ 
          (a) Comparison between the TSHMR from our model and other work. 
             The background density plots show the distributions of modelled galaxies 
             where the color indicates the number density of galaxies.
             TSHMR from literature includes 
             \citealt{Leauthaud2012b} (dashed grey line), 
             \citealt{Budzynski2014} (dashed purple line), 
             \citealt{Patel2015} (solid green line), and 
             \citealt{Kravtsov2018} (dashed teal line). 
             Our median TSHMR is highlighted by grey circles.
          (b) Comparison between the SHMR of central galaxies from our best-fit model 
             and recently published SHMRs at similar redshifts, including
             \citealt{RodriguezPuebla2017} (dot-dashed green line), 
             \citealt{Tinker2017} (dot-dashed brown line), 
             \citealt{Moster2018} (dashed purple line), 
             \citealt{Kravtsov2018} (solid pink line).  
             For the median SHMR: the grey circles show the median \mmax{} at 
             different \mmax{} bins while the grey triangles show the median
             \mvir{} at fixed \mmax{}. 
             Error bars indicate the scatter of \mmax{} or \mvir{} within the bin.
             All comparisons are under the same fiducial assumptions of
             $h$--factor $h=0.7$, Chabrier IMF, and \texttt{FSPS} stellar population 
             models.
          The \texttt{Jupyter} notebook for this figure is available here:
          \href{https://github.com/dr-guangtou/asap/blob/master/note/fig5.ipynb}{\faGithubAlt}
          }
      \label{fig:asap_model_2}
  \end{figure*}

  \begin{figure*}
      \centering 
      \includegraphics[width=13cm]{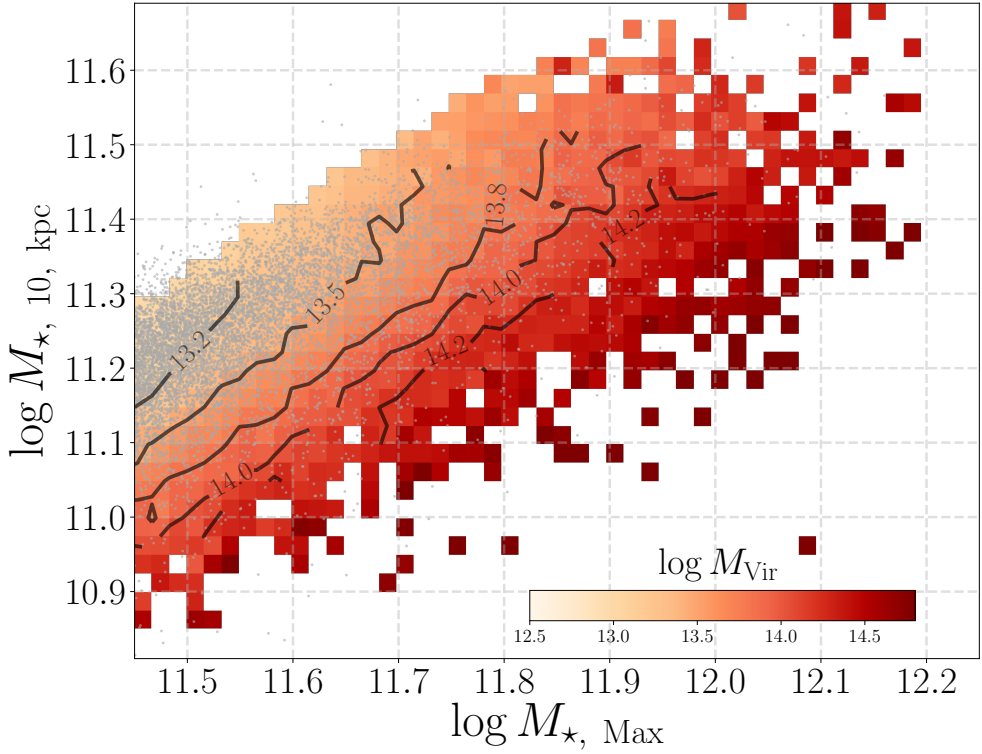}
      \caption{
          Variation of halo mass across the aperture mass plane based on the 
          best-fit \asap{} model for massive central galaxies.
          Grey points in the background show the distribution of HSC galaxies.
          Color of the density plot labels the average \mvir{} from the model. 
          Black contours highlight `iso-\mvir{}' curves over this 2-D plane.
          The \texttt{Jupyter} notebook for this figure is available here:
          \href{https://github.com/dr-guangtou/asap/blob/master/note/fig6.ipynb}{\faGithubAlt}
          }
      \label{fig:asap_model_4}
  \end{figure*}

  \begin{figure*}
      \centering 
      \includegraphics[width=\textwidth]{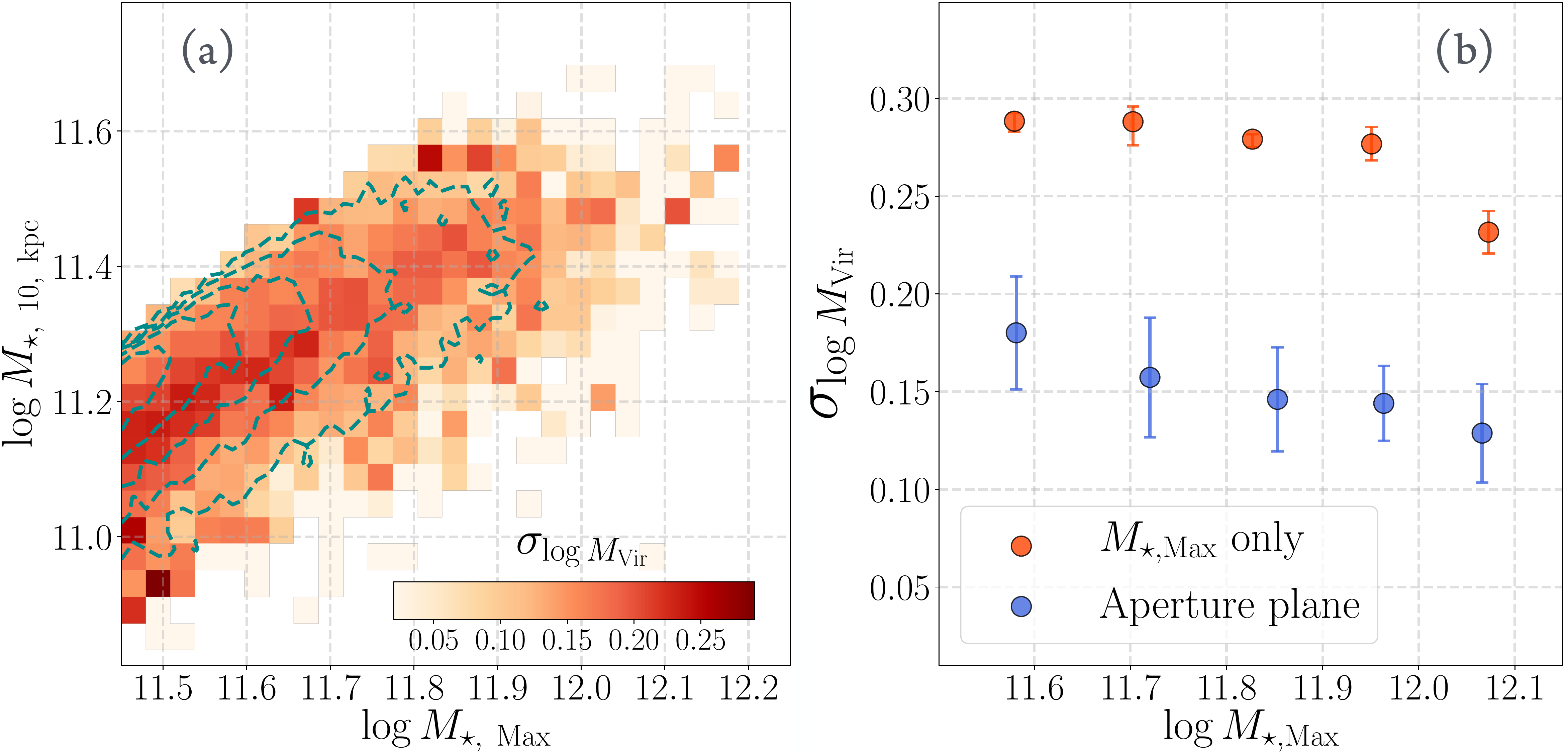}
      \caption{
          (a) Scatter of \logmvir{} across the aperture mass plane. 
          The distribution of HSC galaxies is outlined using contours (grey dashed-line).
          (b) Scatter of \logmvir{} at different \mmax{}.
          Red points correspond to the scatter in different \mmax{} bins. Blue points
          show the mean scatter when information from the aperture mass plane is 
          considered.
          The errors bars are estimated by randomly drawing from the posterior 
          distributions of model parameters. 
          Scatter here includes both an intrinsic component as well as the uncertainty 
          of stellar mass measurements. 
          The \texttt{Jupyter} notebook for this figure is available here:
          \href{https://github.com/dr-guangtou/asap/blob/master/note/fig7.ipynb}{\faGithubAlt}
          }
      \label{fig:asap_model_5}
  \end{figure*}

  \begin{figure*}
      \centering 
      \includegraphics[width=\textwidth]{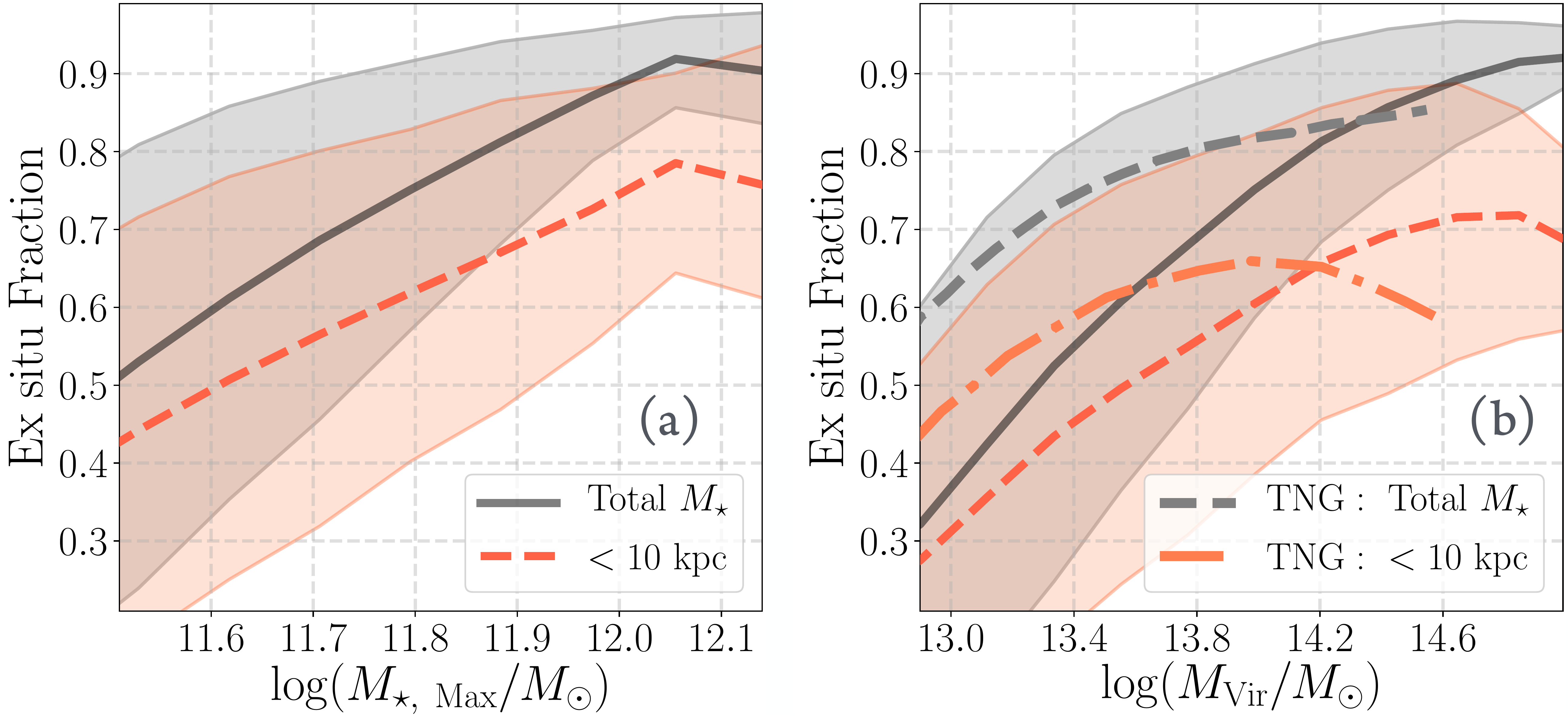}
      \caption{
         (a) The relation between the fraction of \exsitu{} stars and stellar 
          mass (\mmax{}) using the best-fit \asap{} model.
          (b) The same relation between halo mass (\mvir{}). 
          Solid black lines and the corresponding shaded regions are for the \exsitu{}
          fraction in the total stellar mass, while the dashed-red lines and the 
          associated shaded regions indicate the \exsitu{} fraction within 10 kpc. 
          The shaded regions describe the 1$\sigma$ uncertainties. 
          In massive galaxies, \exsitu{} stars dominate the total stellar mass budget     
          \emph{and} the central stellar mass when \logmmax{}$>11.5$ or when 
          \logmvir{}$>13.5$.       
          In (b) we also compare our results with similar relations from the 
          \texttt{TNG300} simulation (see \citealt{Pillepich2017b} for details).
          The \texttt{Jupyter} notebook for this figure is available here:
          \href{https://github.com/dr-guangtou/asap/blob/master/note/fig8.ipynb}{\faGithubAlt}
          }
      \label{fig:asap_model_3}
  \end{figure*}

  \begin{figure*}
      \centering 
      \includegraphics[width=\textwidth]{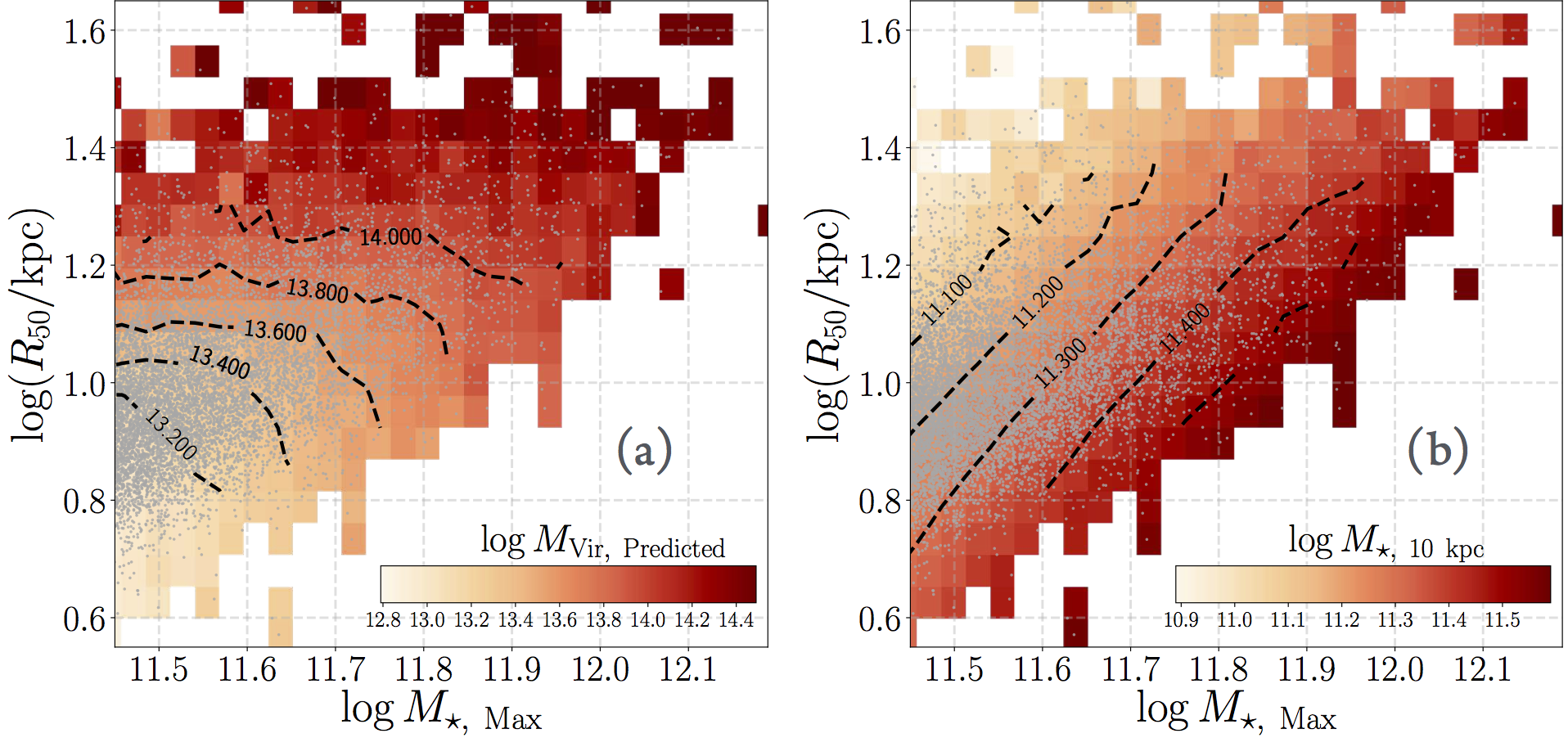}
      \caption{
          Stellar mass (\mmax{})--galaxy size ($R_{50}$) relation color-coded by
          the predicted halo mass (\mvir{}) from the best-fit model (a) and
          the stellar mass within 10 kpc (b).
          For each HSC galaxy, we assign a \mvir{} using the best-fit 
          \mmax{}--\minn{}--\mvir{} relation. 
          The \texttt{Jupyter} notebook for this figure is available here:
          \href{https://github.com/dr-guangtou/asap/blob/master/note/fig9.ipynb}{\faGithubAlt}
          }
      \label{fig:asap_model_6}
  \end{figure*}
          
         
\section{Results}
    \label{sec:result}


\subsection{Fitting our model to the data}
    
    Finally, we combine the likelihood for SMF and \dsigma{} profiles for the model:
    
    \begin{equation}
        \ln \mathcal{L}_{\rm tot} = \ln \mathcal{L}_{\rm SMF} + \sum_{j}^{m} \ln \mathcal{L}_{{\Delta\Sigma}_{j}}
    \end{equation}
    
    \noindent The sum over $j$ is for the $m=12$ aperture mass bins.
    To sample the posterior distributions of model parameters, we choose to use 
    the \href{https://github.com/dfm/emcee}{affine invariant MCMC ensemble sampler 
    \emcee{}} (\citealt{Emcee}).
    We use an ensemble of 256 walkers.
    Following the strategy of the SED fitting code 
    \href{https://github.com/bd-j/prospector}{\texttt{prospector}} 
    (Johnson\etal in prep.), we separate the burn--in stage into three separated 
    rounds, each with 150 steps. 
    We reinitialize the walkers at the end of each round using the current 
    best position of the ensemble and the covariance matrix measured using positions of
    50 per cent walkers.
    This method can effectively remove stalled walkers and helps the chains to 
    converge. 
    After the burn--in stage, we sample another 400 steps to form the posterior 
    distributions of parameters.
    Following \texttt{prospector}, we use the Kullback--Leibler divergence to check
    the convergence of our chains. 
    The trace plot of this model is available here: 
    \href{https://github.com/dr-guangtou/asap/blob/master/note/fig3_default.ipynb}{\faGithubAlt}
    
    We choose weakly informative priors for the seven parameters in our model.
    Following the recommendation by \citet{Gelman2009}, we adopt a Student--$t$ distribution 
    with one degree of freedom as a prior distribution for the slopes in all of the 
    $\log$--linear scaling relations in our model ($a$, $c$, and $A_{\rm exs}$).
    For other parameters ($b$, $d$, $f_{\rm ins}$, and $B_{\rm exs}$), we choose 
    simple top--hat distributions with reasonable boundaries.
    For instance, the the upper limit for \ins{} stars within inner 10 kpc is naturally 1.0. 
    We summarise the prior distributions of all seven parameters in the upper-right table 
    of Figure~\ref{fig:asap_corner}.
    Different choices of prior distributions (e.g. top--hat distributions for all 
    parameters) does not alter key conclusions of this work.
   
    
\subsection{Performance of the best-fit model}
    \label{ssec:best}
    
    Here we summarize the key results from our best-fit model. 
    Figure~\ref{fig:asap_corner} presents the best-fit parameters along with their
    68 per cent confidence intervals. 
    We show the two-dimensional marginalized probability densities of these parameters 
    and the histograms of their marginalized posterior distributions using 
    corner plots\footnote{The corner plot is generated by 
    \href{https://corner.readthedocs.io/en/latest}{\texttt{corner.py}}}. 
    The parameters are well-constrained. 
    The correlations between $a$ and $b$, also between $A_{\rm exs}$ and $B_{\rm exs}$,
    are expected. 
    
    As shown in Figure~\ref{fig:asap_model_1}, the best-fit model is capable of 
    reproducing the observations, including the SMFs for both \mmaxmod{} and \minnmod{},
    and the \dsigma{} profiles in different aperture mass bins. 
    The predicted SMFs of \minnmod{} and \mmaxmod{} are consistent with the observed 
    galaxies at \logmmax{}$>11.6$ within uncertainties. 
    And the predicted SMF of \mmaxmod{} is also consistent with the SMF from the 
    PRIMUS survey at similar redshift range (\citealt{Moustakas2013}) down to
    \logmmax{}$\sim 11.2$ where no observations are included.
    As for the \dsigma{} profiles, the overall goodness-of-fit is excellent, although 
    small mismatches can be found at $>1$ Mpc in a few aperture mass bins (e.g. 
    bin 1, 2, \& 12). 
   
\subsection{Best fit TSMR and SHMR}
    \label{sec:shmr}


\subsubsection{TSHMR}
    \label{ssec:tshmr}
    
    From the best-fit model, we have the TSHMR: 

    \begin{equation}
        \log \mathcal{M}_{\star}^{\mathrm{all}} = 0.602_{-0.006}^{+0.005} \times (\log M_{\rm vir} - 13.5) + 11.846_{-0.003}^{+0.003}
    \end{equation}      
    
    We show the distribution of central galaxies on the \mvir{}--\mallmod{} plane and 
    the median TSHMR in panel (a) of Figure~\ref{fig:asap_model_2}. 
    The best-fit SHMR is indeed very tight. 
    In fact, as indicated by the best-fit $c$ and $d$, the TSHMR has very little 
    scatter.
    As explained earlier, this scatter is supposed to account for both intrinsic 
    scatter and measurement uncertainties. 
    Hence such small scatter is unlikely to be realistic. 
    In Appendix~\ref{app:um_stats}, we show that the current \um{} model has a large 
    scatter of \fgal{} for central galaxies at fixed \mall{}. 
    It seems such scatter alone can account for the intrinsic scatter of SHMR and 
    uncertainties of stellar mass measurements when comparing to observations, 
    practically leaving no room for additional scatter of the TSHMR. 
    Although previous works have also commented on the apparent tightness of the 
    TSHMR (e.g. \citealt{vanderBurg2014, Patel2015, Kravtsov2018}), 
    it remains an open question whether TSHMR scatter is genuinely as tight as indicated 
    by these and our results.
   
    Figure~\ref{fig:asap_model_4} compares the best-fit TSHMR with other
    observational constraints of groups and clusters at similar redshifts.    
    \citet{Leauthaud2012b} constrain the TSHMR of groups in the COSMOS field
    at $0.22 < z < 0.48$. 
    \citet{Budzynski2014} derive the TSHMR for a large sample of low-redshift SDSS 
    groups and clusters using optical richness.  
    \citet{Patel2015} estimate the TSHMR for X-ray groups 
    ($M_{200c}<10^{13.5} M_{\odot}$) in the \textit{Chandra} Deep Field South 
    (CDF-S) field. 
    And \citet{Kravtsov2018} measure the TSHMR for 21 massive $z \sim 0$ clusters 
    using X-ray observations and improved photometric models of massive brightest 
    cluster galaxies (BCGs). 
    The slope of our TSHMR ($0.602\pm0.005$) is shallower than some previous 
    estimates (e.g. $0.89\pm0.14$ in \citealt{Budzynski2014}; 
    $0.84\pm0.10$ in \citealt{Patel2015}; also see
    \citealt{Giodini2009, Lin2012}), but once we convert different 
    TSHMRs into $h=0.7$ with the Chabrier IMF, the overall agreement is good. 
    
    Our TSHMR is constrained by the deepest imaging dataset for a large sample of 
    massive galaxies and high signal-to-noise g--g lensing measurements.
    The best--fit relation is consistent with other observational constraint down to 
    \logmvir{}$\ge 12.5$, which extends below the halo mass range probed by the 
    observed massive galaxies.
    This further suggests that the total stellar mass in a dark matter halo is an 
    excellent proxy of halo mass 
    (although see discussion about the scatter of the TSHMR.).


\subsubsection{SHMR}
    \label{ssec:shmr}
    
    Figure~\ref{fig:asap_model_4} displays the number density 
    distribution of model galaxies (indicated by color) over the \mvir{}--\mmax{} 
    plane. 
    As discussed in \citet{Tinker2017} and \citet{RodriguezPuebla2017}, the 
    non-Gaussian distribution of galaxies along the SHMR causes differences between 
    SHMR when described by the mean \mmax{} at fixed \mvir{} 
    ($\left\langle M_{\star}\right\rangle_{M_{\rm vir}}$; grey circles) and 
    by the mean \mvir{} in bins of \mmax{} 
    ($\left\langle M_{\rm vir}\right\rangle_{M_{\star}}$; grey triangles).
    
    A $\log$-linear fit for 
    $\left\langle M_{\star}\right\rangle_{M_{\rm vir}}$ at 
    $\log M_{\rm vir} \ge 13.0$ yields:

    \begin{equation}
        \log M_{\star}^{\rm max} = 0.36\pm0.01 \times (\log M_{\rm vir} - 13.27) + 11.38\pm0.02
    \end{equation}

    \noindent{}with a scatter of $\sigma_{\log M_{\rm vir}} = 0.23\pm0.01$. 
    The best-fit $\log$--linear relation for 
    $\left\langle M_{\rm vir}\right\rangle_{M_{\star}}$ at 
    $\log M_{\star}^{\rm max} \ge 11.5$ is:
    
    \begin{equation}
        \log M_{\rm vir} = 2.49\pm0.02 \times (\log M_{\star}^{\rm max} - 11.6) + 13.39\pm0.02
    \end{equation}   

    \noindent{}with a scatter of $\sigma_{\log M_{\star}^{\rm max}}=0.22\pm0.01$.
    
    We compare our results with recent constraints of SHMR in the form of
    $\left\langle M_{\star}\right\rangle_{M_{\rm vir}}$\footnote{All SHMR have also 
    been converted to $h=0.7$, Virial halo mass and the Chabrier IMF.}.
    \citet{Tinker2017} estimate the SHMR for massive ($\log M_{\star}\ge 11.4$) 
    CMASS galaxies (e.g, \citealt{Dawson2013}) at $0.4 < z < 0.7$ using 
    clustering measurements. 
    The SHMR from \citet{Kravtsov2018} show here is from an abundance matching
    method based on the SMFs from \citet{Bernardi2013}. 
    The SHMRs of \citet{RodriguezPuebla2017} and \citet{Moster2018} are from 
    two new semi-empirical models that are similar to the \um{} in methodology. 
    
    Recent empirical models (e.g. \citealt{RodriguezPuebla2017, Moster2018, 
    Behroozi2018} have adopted the improved $z\sim 0$ SMF from 
    \citet{Bernardi2013} which uses a better background subtraction. 
    This approach could lead to better agreement with our result using deep HSC 
    images than earlier models that are constrained by local SMFs that underestimate 
    the masses of massive galaxies (e.g. \citealt{Behroozi2013}).
    
    Scatter in the SHMR includes an intrinsic component and uncertainties of 
    stellar mass measurements. 
    Our results agree well with recent constraints when described by 
    $\sigma_{\log M_{\star}}$ at fixed \mvir{}.
    \citet{Tinker2017} find $\sigma_{\log M_{\star}}=0.18_{-0.02}^{+0.01}$ 
    at $\log M_{\star} \ge 11.4$. 
    The \texttt{Emerge} model by \citet{Moster2018} shows a scatter of
    $\sigma_{\log M_{\star}}=0.16$ at high masses. 
    Along with other recent work (e.g. \citealt{Reddick2013, Zu2016}), these
    estimates leave little room for intrinsic scatter in the high mass SHMR 
    ($\sigma_{\log M_{\star}}^{\rm intr}<0.16$). 
    
    In our model, the scatter in the SHMR is a result of the TSHMR and the scatter in
    $f_{\star}^{\rm cen}=M_{\star}^{\rm cen}/M_{\star}^{\rm all}$ as predicted by the 
    \um{} model. 
    Physically, the decreasing scatter of $f_{\star}^{\rm cen}$ may result from the 
    central limit theorem during the complex merging history of massive haloes 
    (e.g. \citealt{Gu2016}). 
    This scatter is discussed in further detail in the following section.

\subsection{Variations of \mvir{} across the \mmax{}--\minn{} plane}
    \label{ssec:mhalo_tren}

    The main goal of our model is to evaluate the \mmax{}--\minn{}--\mvir{} relation. 
    Figure~\ref{fig:asap_model_4} displays variations in \mvir{} across the aperture 
    mass plane. 
    This trend is strongly constrained by the \dsigma{} profiles in different aperture 
    mass bins. 
    The variation is also consistent with the intuition we initially gained from 
    Figure~\ref{fig:asap_model_1}. 
    The amplitude of \dsigma{} increases with \mmax{} but also decreases with \minn{} 
    at fixed \mmax{}. 
    This indicates higher \mvir{} for massive galaxies with more extended stellar 
    envelopes.
    
    The median \mvir{} in each bin is shown on the upper-right corner of each panel. 
    Typically, the range of \mvir{} across the three bins with similar \mmax{} is 
    about 0.15-0.20 dex. 
    But, as shown in panel (a) of Figure~\ref{fig:asap_model_1}, this is caused by
    the choices of \minn{} bins at fixed \mmax{}:
    although they cover very different ranges of \minn{}, the mean \minn{} values for 
    the three bins are not very different due to the distribution of massive galaxies. 
    Right now the choice of mass bins is limited by the required number of galaxies 
    to ensure sufficient \s2n{} of the \dsigma{} profile and uncertainties of 
    stellar mass measurements ($\sim 0.1$ dex). 
    This is not ideal for direct measurement of `local' \mvir{} across the aperture
    mass plane and is precisely why we choose to use the forward modelling approach 
    by simultaneously considering twelve \dsigma{} profiles and two SMFs so that we 
    can still use the best-fit model to explore the \mvir{} trend in more detail. 
        
    The iso-\mvir{} curves on Figure~\ref{fig:asap_model_5} run almost parallel to 
    the \mmax{}--\minn{} relation, resulting in a considerable range of \mvir{} 
    ($> 0.7$ dex) in the vertical direction at fixed \mmax{}. 
    This range is not surprising, however, given the range of \mvir{} seen on the 
    SHMR at fixed \mmax{} (e.g. see Figure~\ref{fig:asap_model_2}; also see Figure~9 
    in \citet{Tinker2017}). 
    What is surprising, however, and this is one of the main results of this paper, 
    is that a large fraction of the scatter can be accounted for by structural 
    variations in massive galaxies. 
    In other terms, the scatter in \mvir{} in greatly reduced in the aperture mass 
    plane compared to the SHMR. 
    Figure~\ref{fig:asap_model_5} displays the scatter of \mvir{} 
    ($\sigma_{\log M_{\rm vir}}$) across the aperture mass plane. 
    Among regions occupied by most massive galaxies (indicated by the contours), 
    the typical scatter is only of order 0.15 dex.

	Figure \ref{fig:asap_model_5} suggests that the combination of \mmax{}--\minn{} 
	predicts \mvir{} better than \mmax{} alone. 
	For instance, a simple random forest regressor\footnote{
	e.g. The \texttt{RandomForestRegressor} from \texttt{scikit-learn} package.
	Random forest is a flexible machine learning algorithm that uses a combinations 
	of multiple decision trees to make predictions based on the data.} can provide 
	an accurate description of the \mmax{}--\minn{}--\mvir{} 3-D space and can be used 
	to predict \mvir{} (see Appendix~\ref{app:mvir}). 
    However, with random forest there is a risk of overfitting, and the results are 
    not intuitive. 
    We therefore also fit the \mmax{}--\minn{}--\mvir{} plane using the robust linear 
    regression algorithm \href{https://pypi.org/project/ltsfit}{\texttt{LtsFit}}
    (\citealt{Cappellari2014}).
    The best-fit relation is:
   
    \begin{multline}
        \log M_{\rm vir} = 3.26\pm0.02 \times (\log M_{\star}^{\rm max} - 11.72) \\
        \mathrm{\ }-2.46\pm0.03 \times (\log M_{\star}^{10} - 11.34) + 13.69\pm0.01
    \end{multline}

    \noindent{}with a scatter of $\sigma \log M_{\rm vir}=0.16\pm0.01$.    
    As shown in Appendix~\ref{app:mvir}, this simple relation is also capable of 
    predicting \mvir{} with reasonable precision and a smaller scatter than the SHMR.
    We further discuss predictive capabilities in Section~\ref{sec:clash}. 
    
\subsection{\emph{In situ} and \exsitu{} fractions}
    \label{ssec:}
    
    Since the version of \um{} used here predicts the \mstar{} of the \ins{} and \exs{} 
    components, our model can be used to shed light on the statistical behaviors of 
    these two components. 
    The best-fit model suggests that $67\pm 1$ per cent of \ins{} stars can be found 
    within 10 kpc, while the fraction of \exs{} stars within 10 kpc slowly decreases 
    with \mvir{} due to the increasingly extended distribution of the \exs{} component. 
    At \mvir{}$=10^{13} M_{\odot}$, about half of the \exs{} stars lie inside 10 kpc
    according to the best-fit model. 
    This fraction decreases to $\sim 30$ per cent for a \mvir{}$=10^{14} M_{\odot}$ 
    halo. 
    
    Focusing on the \exs{} component, we show how the fraction of \exs{} stars 
    changes with \mmax{} and \mvir{} in Figure~\ref{fig:asap_model_3}.
    In agreement with results from recent hydro-simulations 
    (e.g. \citealt{RodriguezGomez2016, Qu2017, Pillepich2017}), the \exs{} fraction 
    increases with both stellar and halo mass, and it remains the dominant stellar 
    component in massive galaxies. 
    Remarkably, this is not just the case for the galaxy as a whole, but it is even true on 10 kpc 
    scale for these massive galaxies. 
    For central galaxies with \logmmax{}$>11.5$ and in haloes with 
    \logmvir{}$>13.5$, the average \exs{} fraction at $r<$10 kpc is $>50$ per cent. 
    
    In Figure~\ref{fig:asap_model_3}, we also compare the trends of the \exs{} 
    fraction with halo mass with results from the \href{http://www.tng-project.org}{
    \texttt{IllustrisTNG}} simulation
    (\citealt{Pillepich2017}; \texttt{TNG} hereafter).
    We find reasonable qualitative agreement between our model and the \texttt{TNG} 
    simulation; differences in detail are to be expected, given the different methods used 
    for measuring \mstar{} (Ardilla et al in prep), e.g., 
    for the \exs{} fraction within 10 kpc, \citet{Pillepich2017} use a 3-D sphere 
    while we use a 2-D elliptical aperture.
        
    The dominant role of \exs{} stars at the centers of massive galaxies has been 
    discussed by \citet{Cooper2013} using a particle-tagging method and by 
    \citealt{RodriguezGomez2016} using the \texttt{Illustris} simulation. 
    It is likely that these \exs{} stars originate from major mergers 
    that happened at high-$z$. 
    This is directly related to the current definition of the \ins{} and \exs{} 
    components. We discuss this further in Section~\ref{sec:fexsitu}.

\subsection{Relation between \mhalo{} and galaxy size}
    \label{ssec:size}

    Figure~\ref{fig:asap_model_6} shows variations in the mean \mvir{} across the 
    mass--size relation.
    Here we use \mmax{} and a non-circulized half-light radius measured using a 1-D 
    stellar mass curve of growth. 
    For each massive galaxy, we assign a \mvir{} using the best-fit 
    \mmax{}--\minn{}--\mvir{} relation derived above. 
    We do not attempt to remove satellite galaxies. 
    
    In \citetalias{Huang2018b}, we showed that massive central galaxies in halos of 
    different mass exhibit a distinct mass--size relation. 
    Panel (b) of Figure~\ref{fig:asap_model_6} presents more sophisticated constraints 
    on this `environmental' dependence of mass--size relation: \mvir{} varies
    systematically across this plane, and the iso-\mvir{} curves here are almost 
    perpendicular to the mass--size relation. 
    {\em At fixed \mmax{}, galaxies with larger size tend to live in more massive 
    haloes. }
    This suggests that the sizes of massive galaxies also carry clues about their 
    dark matter halo mass, as discussed in \citet{Kravtsov2013}. 
    However, as discussed in \citetalias{Huang2018b}, the measurement of 
    `galaxy size' often depends on the assumed photometric model and data quality.
    Therefore, we prefer to build our empirical model based on a more straightforward 
    aperture mass plane instead of the mass--size relation. 
    

  \begin{figure}
      \centering 
      \includegraphics[width=\columnwidth]{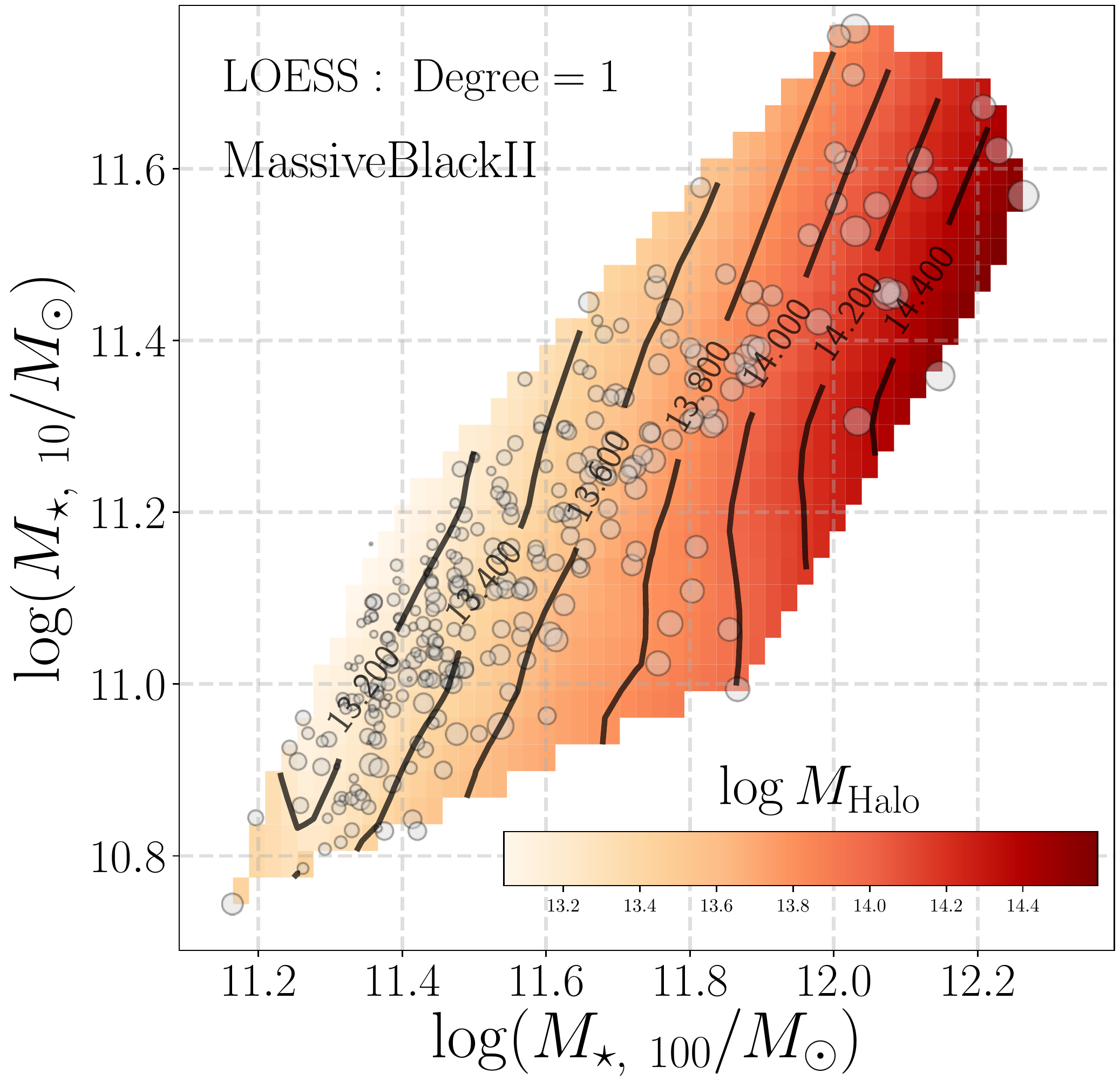}
      \caption{
          The \mtot{} and \minn{} of massive galaxies from the \texttt{MassiveBlackII} 
          simulation using randomly projected 2-D stellar mass distributions. 
          The plot shows their distributions over the aperture mass plane color-coded 
          by halo mass. 
          The density plot shows the halo mass trend recovered by the LOESS smoothing 
          method.
          The \texttt{Jupyter} notebook for this figure is available here:
          \href{https://github.com/dr-guangtou/asap/blob/master/note/fig10.ipynb}{\faGithubAlt}
          }
      \label{fig:mblack}
  \end{figure}

  \begin{figure*}
      \centering 
      \includegraphics[width=\textwidth]{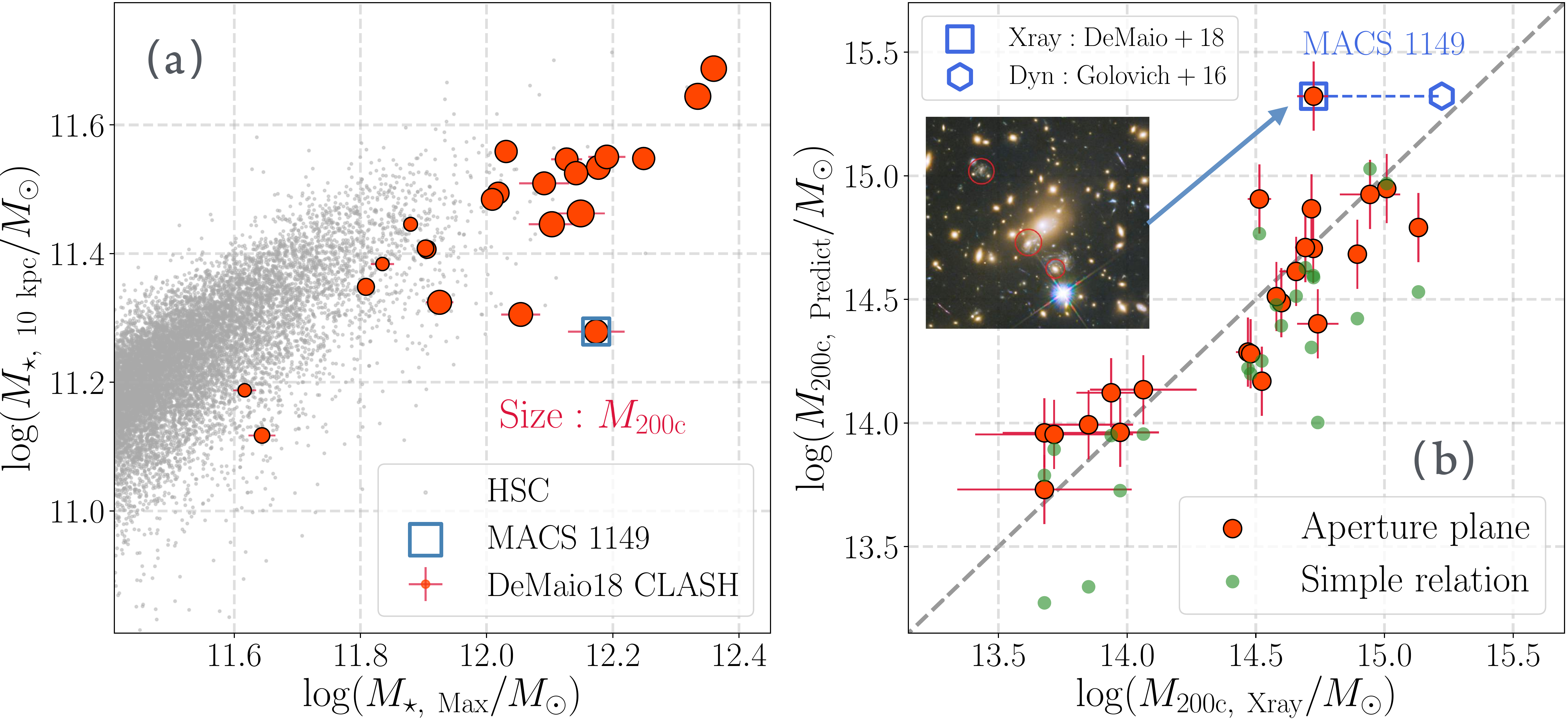}
      \caption{
          (a) Distribution of CLASH BCGs in the aperture mass plane. 
              The stellar masses are based on 
              Hubble Space Telescope (HST) observations (\citealt{DeMaio2018}; 
              red points), 
              where the symbol size indicates halo mass ($M_{200c}$) estimated from 
              X-ray observations. 
          (b) Comparison between our predicted $M_{200c}$ values to those based on X-ray 
              observations.
              Green points are predictions based on the \mmax{}--\mhalo{} relation,
              whereas red points are based on the best-fit aperture mass plane.
              The dashed line shows the one-to-one relation. 
              The prediction using the aperture mass plane shows a tighter relation 
              compared to X-ray masses. 
              The outlier system, MACS1149, is highlighted using a blue box. 
              We also show the colored HST image of the BCG of MACS1149. 
              The blue hexagon shows the value of $M_{200c}$ as measured via dynamics 
              \citep[][]{Golovich2016}.
          The \texttt{Jupyter} notebook for this figure is available here:
          \href{https://github.com/dr-guangtou/asap/blob/master/note/fig11.ipynb}{\faGithubAlt} 
          }
      \label{fig:clash}
  \end{figure*}

  \begin{figure*}
      \centering 
      \includegraphics[width=\textwidth]{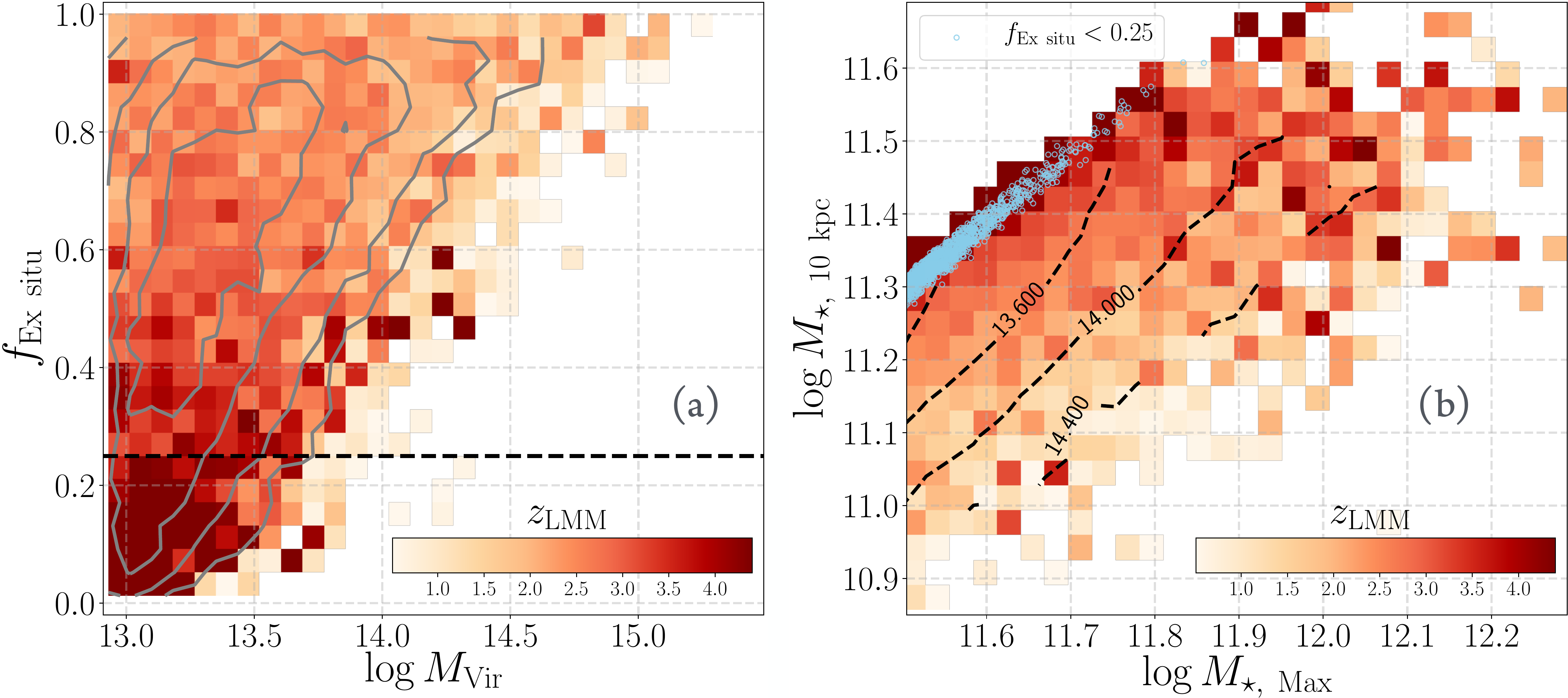}
      \caption{
          (a) Relation between the halo mass (\mvir{}) and the fraction 
              of \exs{} stars for massive galaxies (\logmmax{}$\ge 11.5$) 
              color-coded by the redshift of last major-merger ($z_{\rm LMM}$; 
              halo mass ratio larger than 1:3) in the \um{} model. 
              The grey contours outline the number density distribution of 
              model galaxies. 
              A horizontal dashed line highlights the 25 per cent limit that is used 
              to define massive galaxies with low \exs{} fraction. 
          (b) The aperture mass plane for massive galaxies from the best-fit 
              model color-coded by $z_{\rm LMM}$.
              Grey contours here indicate the iso-\mvir{} lines. 
              The small population of massive model galaxies with $f_{\rm exs} < 0.25$
              is highlighted with light--blue dots.
          The \texttt{Jupyter} notebook for this figure is available here:
          \href{https://github.com/dr-guangtou/asap/blob/master/note/fig12.ipynb}{\faGithubAlt}
          }
      \label{fig:zlmm}
  \end{figure*}


\section{Discussion}
    \label{sec:discussion}


\subsection{Comparison with hydrodynamic simulations}
    \label{sec:hydro}
    
    From HSC g--g lensing measurement and our best-fit model, we find that the halo 
    masses of massive galaxies vary systematically across the aperture mass plane. 
    This reveals a clear connection between the distribution of stars within massive 
    galaxies, and halo mass. 
    We now investigate if such correlations are also predicted by hydrodynamic 
    simulations of galaxy evolution.
    
    
    We compare the observed \mmax{}--\minn{}--\mhalo{} relation with the relations of 
    massive galaxies from the \href{http://mbii.phys.cmu.edu}{\texttt{MassiveBlackII}} 
    simulation (e.g. \citealt{Khandai2015, Tenneti2015}).
    \texttt{MassiveBlackII} is a state-of-the-art, large-volume 
    (100$h^{-1}$ Mpc box size; $1792^3$ gas particles),
    high-resolution cosmological simulation using \texttt{p-Gadget} 
    (\citealt{Springel2005}).
    It includes a sophisticated treatment of complex baryonic physics
    (e.g. star formation in a multiphase interstellar medium, black hole accretion 
    and feedback, and radiative cooling and heating processes). 
    For additional information about the physical details and general performance of 
    the \texttt{MassiveBlackII} simulation, please refer to \citet{Khandai2015}.
    
    We select 291 massive galaxies with $\log (M_{\star}/M_{\odot}) \ge 11.4$ from 
    the \texttt{MassiveBlackII} simulation and generate randomly projected 2--D 
    stellar mass maps with a 2 kpc pixel resolution and 350 kpc image size. 
    Then we treat them as real data and measure their aperture masses using the 
    same method for HSC massive galaxies (Ardila et al. in prep.).
    We choose to use 10 and 100 kpc elliptical apertures here. 
    In Figure~\ref{fig:mblack}, we show the trend of halo mass across this aperture 
    mass plane recovered by the locally weighted regression (LOESS) method 
    (\citealt{Cleveland1988, Cappellari2013b})\footnote{We use the Python 
    implementation of \href{https://pypi.org/project/loess/}{2-D LOESS} by 
    Michelle Cappellari.}
    The trend is qualitatively similar to our results, while the slope of the 
    iso--\mhalo{} curves appears to be steeper than those of our best--fit model.  
        

    Currently, comparison with simulation is limited by the volume of high-resolution 
    hydro-simulations and their capabilities to reproduce realistic massive galaxies.
    The SMF of massive galaxies using \mtot{} and the stellar mass density profiles of 
    massive galaxies \texttt{MassiveBlackII} do show differences compared with the 
    HSC observations (see Ardilla\etal{}~in prep.). 
    Nonetheless, we consider this to be a valuable test and we will further investigate 
    the robustness of this trend using other hydro simulations in future work.
    
        

\subsection{Prediction of halo mass}
    \label{sec:clash}
    
    Our \asap{} model suggests that, by including information about the stellar mass 
    distribution (e.g. two-aperture stellar masses), one can build better proxies of 
    halo mass. 
    We test this potential by predicting the halo masses of massive clusters
    from the \href{https://archive.stsci.edu/prepds/clash/}{Cluster Lensing And 
    Supernova survey with Hubble} (CLASH) clusters (e.g. \citealt{Postman2012b})
    using only the photometry of their brightest cluster galaxies. 
    
    \citet{DeMaio2018} conducted a careful study of the BCG$+$ICL of 23 
    CLASH clusters 
    ($0.3 < z < 0.9$; $3\times 10^{13} < M_{500c}/M_{\odot} < 9\times10^{14}$)
    using multiband, high-resolution HST Wide Field Camera 3 (WFC3) images. 
    These authors derive surface brightness and color profiles of these BCG$+$ICL to 
    $r>$100 kpc, along with stellar mass within 10 and 100 \emph{circular} 
    apertures using SED fitting. 
    We ignore the differences caused by circular and elliptical apertures here and 
    increase their 100 kpc aperture mass by $+0.05$ dex to simulate our \mmax{} 
    measurement (see \citetalias{Huang2018b}).
    After converting their aperture masses to the same cosmology and stellar 
    population model\footnote{\citet{DeMaio2018} uses the BC03 stellar population 
    model. 
    Based on tests from \citetalias{Huang2018b}, we add a $+0.1$-dex empirical 
    correction to the Flexible Stellar Population Synthesis (FSPS) mode used in 
    this work.} used here, we predict the \mhalo{} of these BCGs using our best-fit 
    model.
    The CLASH sample includes mostly very massive clusters that host BCGs that are 
    on average more massive than the HSC sample (panel (a) of Figure~\ref{fig:clash}).

    Figure~\ref{fig:clash} shows halo masses predicted both by the average 
    \mmax{}--\mvir{} relation shown in Figure~\ref{fig:asap_model_2} (green dots) and 
    by the best-fit \mmax{}--\minn{}--\mvir{} relation (red circle). 
    In \citet{DeMaio2018}, halo mass is measured using X-ray observations (e.g. 
    \citealt{Vikhlinin2009}) and is defined as $M_{500c}$.  
    Using empirical relations from \citep{Diemer2013, Diemer2015} and the 
    \texttt{Colossus} Python package (\citealt{Diemer2017}; code available here:
    \href{https://bitbucket.org/bdiemer/colossus}{\faBitbucket}),
    we convert both the $M_{500c}$ in \citet{DeMaio2018} and the \mhalo{} from our 
    model to $M_{200c}$. 
    It is encouraging to see that the predicted halo mass values show good 
    consistency with those based on X-rays. 
    The values predicted using \mmax{} alone show larger scatter compared to 
    the X-ray mass estimates. 
    This provides further evidence for one of the key findings of the present work: 
    two-aperture stellar masses can be used to build better proxies of halo mass 
    relative to models using \mmax{} alone. 
    
    There is one BCG that shows a large offset (highlighted in both panels of 
    Figure~\ref{fig:clash}) from the mean relations. 
    The BCG belongs to the famous cluster MACS~1149$+$22 at $z=0.544$ (see the inset
    picture) that gifted us multiple images of a highly magnified supernova (e.g.
    \citealt{Kelly2015}) and a $z \sim 9.1$ galaxy (e.g. \citealt{Hashimoto2018}).
    The region around the BCG is complex and partially overlaps with an image of a 
    background star-forming galaxy.
    We suspect that  the accuracy of photometry and \m2l{} estimation are affected by 
    the complexity of extracting photometry for this system.
    Moreover, it is possible that the X-ray gas underestimates the halo mass 
    due to non-thermal pressure support or projection effect 
    (e.g. \citealt{Evrard1990, Nagai2007, Mahdavi2008}). 
    \citet{Golovich2016} estimate the halo mass of MACS~1149$+$22 using
    dynamics of cluster members. 
    The dynamics-based $M_{200c}$ is higher than the X-ray value and is closer
    to our prediction. 
    
    At the same time, we acknowledge that different methods sometimes lead to 
    systematically different measurements of $M_{200c}$.
    For example, it is known that $M_{200c}$ calibrated using weak lensing is often larger compared to X--ray based masses for massive clusters 
    (e.g. \citealt{Simet2017}). 
    Indeed, weak lensing measurements of some massive CLASH clusters in \citet{DeMaio2018}
    result in noticeably more massive $M_{200c}$ than those derived from X--ray's
    (see \citealt{Umetsu2014}). These types of offsets are beyond the scope of this work, but are worth 
    investigating in the future to further improve our predictions of halo mass using aperture 
    stellar masses.
     
\subsection{The fraction of \exsitu{} stars in massive galaxies}
    \label{sec:fexsitu}
    
    
    Figure~\ref{fig:asap_model_3} shows that the \exs{} fraction predicted by our model 
    and its relation with both stellar and halo mass are reasonable and are 
    qualitatively consistent with hydro-simulation (e.g. \texttt{TNG300}). 
    We now discuss two points in Figure~\ref{fig:asap_model_3} of noteworthy interest.
    
    First, the large scatter in \exs{} fractions at fixed stellar or halo mass
    suggests a small population of massive galaxies with low \exs{} fractions 
    ($< 25$ per cent). 
    This special population could experience fewer mergers (especially major mergers) 
    and is an interesting sample to study in greater detail.
    
    On panel (a) in Figure~\ref{fig:zlmm}, we color-code the 
    \mvir{}--$f_{\rm exs}$ relation for massive galaxies (\logmmax{}$\ge 11.5$) in our 
    best-fit model using the redshift of the last major halo merger (halo mass ratio 
    larger than 1:3) extracted from the merger trees of \smdpl{} haloes.
    We find that massive galaxies with low \exs{} fraction tend to live in relatively
    low mass haloes and have not experienced major--mergers in the last 10 Gyrs, 
    putting them among the oldest massive haloes in the universe.
    This small ($\sim 9$ per cent of massive galaxies with $11.5 <$\logmmax{}$\ge 11.7$) 
    population locates exclusively on the upper-edge of the aperture mass plane 
    (panel (b) of Figure~\ref{fig:zlmm}).
    Such a special location suggests that they are much more compact than the similarly 
    massive ones with a richer merging history. 
    If haloes with such a unique assembly history are not artifacts of the \um{} model, 
    they could be very useful for studying galaxy assembly bias (e.g. 
    \citealt{CooperGallazzi2010, Wang2013, Zentner2014, Lin2016}) or for 
    providing a template of the distribution of \ins{} stars in massive haloes. 
    The galaxies discussed here would be somewhat different in nature than `relic' 
    galaxies\footnote{Typically defined as nearby compact quiescent galaxies with 
    stellar mass and effective radius similar to the quiescent galaxies at high--redshift.}
    (e.g. \citealt{Trujillo2014, Arriba2016, Yildirim2017, FerreMateu2017}): 
    the population under discussed here are more massive than  typical relic 
    galaxies, are larger in size, and are unaffected by stripping as this sample is 
    predominantly centrals. 
    
    Second, both the \asap{} model and hydro-simulations predict a high \exsitu{}
    fraction in the inner regions of massive galaxies. 
    This is easy to understand given the current definition of \exsitu{} stars. 
    This is commonly defined as all the stars that are formed outside the halo of the 
    \emph{main progenitor}. 
    The \exsitu{} component therefore includes stars that were accreted from major mergers 
    at very high redshift (e.g. $z > 2$). 
    Although this is a straightforward definition, it may not be the best choice to 
    relate to observational studies of the stellar assembly history of massive galaxies 
    for two reasons. 
    First, it makes the \exs{} component heterogeneous since \exs{} stars can be 
    formed at very different epochs and in haloes with a wide range of \mvir{}. 
    Second, it is hard to separate the \ins{} and \exs{} stars in the inner regions of 
    massive galaxies because stars in both components are assembled at a very early time 
    and share similar stellar population and kinematic properties. 
    Although it is beyond the scope of this work, we argue that it may be worth 
    considering alternative and potentially more instructive decompositions for 
    massive galaxies.


\section{Summary and Conclusions}
    \label{sec:summary}
    
    Using data from the HSC survey, we perform careful aperture mass and weak lensing 
    measurements for a sample of $\sim 3200$ \logmmax$>11.6$ super massive galaxies. 
    Using weak lensing, we reveal a tight connection between the stellar mass 
    distribution of super massive central galaxies and their total dark matter halo 
    mass. 
    {\em At fixed `total' stellar mass (\mmax{}), massive galaxies with more extended mass 
    distributions tend to live in more massive dark matter haloes. }
    This provides a new an independent confirmation, backed by direct weak lensing 
    measurements, of the results from \citetalias{Huang2018c} that \mvir{} varies 
    systematically over the aperture \mstar{} plane. 
    
    To model both the weak lensing and the aperture stellar mass functions, we build 
    a full forward model based on a special version of the semi-empirical model \um{} 
    and the \smdpl{} simulation. 
    \um{} leverages the ability of high-resolution simulations to identify and track 
    the full merger history of dark matter halos; using \um{} as the bedrock of our model 
    allows us to study the co-evolution of massive galaxies and their halos. 
    We augment the baseline \um{} model with two prescriptions that allow us to fit HSC 
    data and predict aperture masses. 
    Our model make the two following assumptions. 
    We assume (1) a tight correlation between halo mass and the mass of its entire 
    stellar content (TSHMR) and (2) a certain fraction of the \ins{} and \exs{} stars 
    locate within the inner 10 kpc of massive galaxies. 
    In our model, the well-studied SHMR and its scatter emerge from the TSHMR. 
    We show that this model provides an excellent description of the observed SMFs for 
    \mmax{} and \minn{}, as well as the \dsigma{} profiles in a series of 
    \mmax{}--\minn{} bins. 
    
    The main conclusions from the current best-fit model include the following:

    
    \begin{itemize}
    
        \item \mvir{} varies systematically over the aperture mass plane. 
              The iso-\mhost{} curves run almost parallel with the direction of the
              \mmax{}-\minn{} relation. 
              The model confirms that at fixed \mmax{}, galaxies with more extended 
              stellar mass distributions (lower \minn{} or larger size) live in more 
              massive dark matter haloes.
              It also shows that scatter in \mvir{} at either fixed \mmax{} or \minn{}
              is quite large.
        
        \item The above trends can be summarized into a simple \mmax{}--\minn{}--\mvir{}
              relation that provides a tighter connection with halo mass than \mmax{} 
              alone. 
              
        \item The usage of two aperture masses can help reduce the scatter in halo 
              mass at fixed total stellar mass. 
              While the standard SHMR in the form of 
              $\left\langle M_{\rm vir}\right\rangle_{M_{\star}}$ typically shows scatter 
              in halo mass of $\sim 0.25$ dex at $11.5 < \log M_{\star}^{\rm max}$, this 
              scatter can be reduced to the $\sim 0.15$ dex level by utilizing our results 
              based on \mmax{}--\minn{}--\mvir{} scaling relation. 
        
        \item Our model predicts that the \exsitu{} fraction increases with both the 
              stellar and halo mass; and it shows that the \exs{} component even 
              dominates the inner 10 kpc of massive galaxies. 
              These predictions are consistent with results from the \texttt{TNG} 
              simulations.
        
    \end{itemize}

    Our results strongly suggest that information about the assembly history of 
    massive dark matter haloes is encoded in the stellar mass distributions of 
    their massive central galaxies. 
    This opens a new window for studying the assembly histories of group- and cluster-mass haloes 
    by using carefully derived proxies based on massive galaxy profiles.



\section*{Acknowledgements}

  The authors would like to thank Frank van den Bosch, Sandra Faber, Joel Primack
  for useful discussions and suggestions.
  
  This material is based upon work supported by the National Science Foundation under 
  Grant No. 1714610. 
  
  The authors acknowledge support from the Kavli Institute for Theoretical Physics.
  This research was also supported in part by National Science Foundation under Grant 
  No. NSF PHY11-25915 and Grant No. NSF PHY17-48958
  
  AL acknowledges support from the David and Lucille Packard foundation, and from the 
  Alfred .P Sloan foundation.

  The Hyper Suprime-Cam (HSC) collaboration includes the astronomical communities of 
  Japan and Taiwan, and Princeton University.  The HSC instrumentation and software were
  developed by National Astronomical Observatory of Japan (NAOJ), Kavli Institute
  for the Physics and Mathematics of the Universe (Kavli IPMU), University of Tokyo,
  High Energy Accelerator Research Organization (KEK), Academia Sinica Institute
  for Astronomy and Astrophysics in Taiwan (ASIAA), and Princeton University.  
  Funding was contributed by the FIRST program from Japanese Cabinet Office,  Ministry 
  of Education, Culture, Sports, Science and Technology (MEXT), Japan Society for 
  the Promotion of Science (JSPS), Japan Science and Technology Agency (JST), Toray 
  Science Foundation, NAOJ, Kavli IPMU, KEK, ASIAA, and Princeton University.
   
  Funding for SDSS-III has been provided by Alfred P. Sloan Foundation, the 
  Participating Institutions, National Science Foundation, and U.S. Department of
  Energy. The SDSS-III website is http://www.sdss3.org.  SDSS-III is managed by the
  Astrophysical Research Consortium for the Participating Institutions of the SDSS-III
  Collaboration, including University of Arizona, the Brazilian Participation Group,
  Brookhaven National Laboratory, University of Cambridge, University of Florida, the
  French Participation Group, the German Participation Group, Instituto de Astrofisica
  de Canarias, the Michigan State/Notre Dame/JINA Participation Group, Johns Hopkins
  University, Lawrence Berkeley National Laboratory, Max Planck Institute for
  Astrophysics, New Mexico State University, New York University, Ohio State University,
  Pennsylvania State University, University of Portsmouth, Princeton University, the
  Spanish Participation Group, University of Tokyo, University of Utah, Vanderbilt
  University, University of Virginia, University of Washington, and Yale University.
  
  The Pan-STARRS1 surveys (PS1) have been made possible through contributions of  
  Institute for Astronomy; University of Hawaii; the Pan-STARRS Project Office; 
  the Max-Planck Society and its participating institutes: the Max Planck Institute 
  for Astronomy, Heidelberg, and the Max Planck Institute for Extraterrestrial Physics, 
  Garching; Johns Hopkins University; Durham University; University of Edinburgh; 
  Queen's University Belfast; Harvard-Smithsonian Center for Astrophysics; Las 
  Cumbres Observatory Global Telescope Network Incorporated; National Central 
  University of Taiwan; Space Telescope Science Institute; National Aeronautics 
  and Space Administration under Grant No. NNX08AR22G issued through the Planetary 
  Science Division of the NASA Science Mission Directorate; National Science 
  Foundation under Grant No. AST-1238877; University of Maryland, and Eotvos 
  Lorand University. 
  
  This research makes use of software developed for the Large Synoptic Survey 
  Telescope. We thank the LSST project for making their code available as free 
  software at http://dm.lsstcorp.org.
  
  The CosmoSim database used in this research is a service by the Leibniz-Institute for 
  Astrophysics Potsdam (AIP).
  The MultiDark database was developed in cooperation with the Spanish MultiDark 
  Consolider Project CSD2009-00064.
  
  This research made use of:
  \href{http://www.stsci.edu/institute/software_hardware/pyraf/stsci_python}{\texttt{STSCI\_PYTHON}},
      a general astronomical data analysis infrastructure in Python. 
      \texttt{STSCI\_PYTHON} is a product of the Space Telescope Science Institute, 
      which is operated for NASA by Association of Universities for Research 
      in Astronomy (AURA);
  \href{http://www.scipy.org/}{\texttt{SciPy}},
      an open source scientific tool for Python (\citealt{SciPy});
  \href{http://www.numpy.org/}{\texttt{NumPy}}, 
      a fundamental package for scientific computing with Python (\citealt{NumPy});
  \href{http://matplotlib.org/}{\texttt{Matplotlib}}, 
      a 2-D plotting library for Python (\citealt{Matplotlib});
  \href{http://www.astropy.org/}{\texttt{Astropy}}, a community-developed 
      core Python package for astronomy (\citealt{AstroPy}); 
  \href{http://scikit-learn.org/stable/index.html}{\texttt{scikit-learn}},
      a machine-learning library in Python (\citealt{scikit-learn}); 
  \href{https://ipython.org}{\texttt{IPython}}, 
      an interactive computing system for Python (\citealt{IPython});
  \href{https://github.com/kbarbary/sep}{\texttt{sep}} 
      Source Extraction and Photometry in Python (\citealt{PythonSEP});
  \href{https://jiffyclub.github.io/palettable/}{\texttt{palettable}},
      colour palettes for Python;
  \href{http://dan.iel.fm/emcee/current/}{\texttt{emcee}}, 
      Seriously Kick-Ass MCMC in Python;
  \href{http://bdiemer.bitbucket.org/}{\texttt{Colossus}}, 
      COsmology, haLO and large-Scale StrUcture toolS (\citealt{Colossus}).

\bibliographystyle{mnras}
\bibliography{asap}

\appendix

\section{\mhalo{} trends of key predictions in the \um{} model} 
	\label{app:um_stats} 
    
    As explained in \S \ref{sec:um_asap}, besides \mhalo{}, \asap{} model also 
    relies on three key predictions from the special version of \um{} model used in 
    this work:
    
    \begin{enumerate}
    
        \item $\delta_{\rm cen}$: the ratio between the stellar mass of central galaxy 
            and the total stellar mass within the halo. 
            This parameter reflects the ``dominance'' of central galaxy in the halo.
            It is determined by the complex merger history of both the halo and the 
            central galaxy.
        \item $\delta_{\rm ins}$ and $\delta_{\rm exs}$: the fractions of stellar mass
            in the \ins{} and \exs{} components for each galaxy.
            They are determined by both the star--formation and mass-assembly history 
            of each galaxy in the halo. 
        
    \end{enumerate}

\begin{figure*}
      \centering 
      \includegraphics[width=\textwidth]{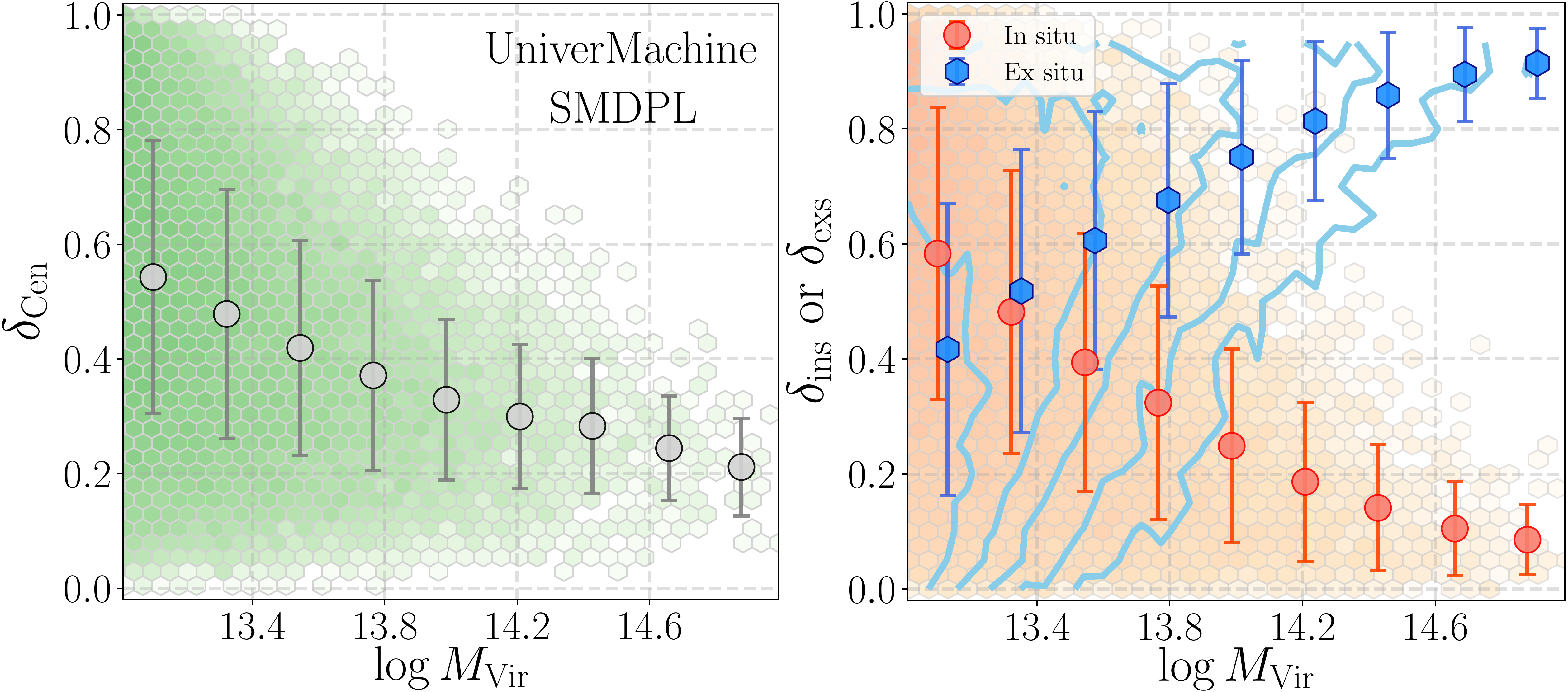}
      \caption{
          Distributions of \mstar{} fraction of central galaxy 
          ($M_{\star}^{\rm cen}/M_{\star}^{\rm all}$) and its dependence on \mvir{}. 
          Color shows the number density of galaxies in $\log$--scale. 
          The median central mass fractions in a series of \mvir{} bins are highlighted 
          using grey circles along with the 1-$\sigma$ scatter in each bin. 
          The shaded region on the left side is for the \mvir{} range ignored in 
          this work.
          The \texttt{Jupyter} notebook for this figure is available here:
          \href{https://github.com/dr-guangtou/asap/blob/master/note/figA1.ipynb}{\faGithubAlt}
          }
      \label{fig:um_stats}
  \end{figure*}

    In Fig~\ref{fig:um_stats}, we visualize their relationships with \mhalo{}. 
    On the left side, we show that $\delta_{\rm cen}$ slowly decrease with \mhalo{}, 
    suggesting that central galaxies become less dominated in more massive halos.
    Meanwhile, this relation has a significant scatter, especially for halos with 
    \logmh{}$\le 14.0$. 
    The scatter of $\delta_{\rm cen}$ at fixed \mhalo{} dominates the scatter of 
    the SHMR predicted by our \asap{} model, although it is yet to learn whether 
    such large scatter is realistic. 
    
    On the right side of Fig~\ref{fig:um_stats}, we demonstrate that the \ins{} (\exs{}) 
    mass fraction rapidly decreases (increases) with \mhalo{}, which is consistent with 
    results from recent hydro--dynamic simulations.
    
\section{Performance of \mvir{} estimators} 
	\label{app:mvir} 
    
    As mentioned in \S\ref{ssec:mhalo_tren}, we attempts to assign \mhalo{} to 
    massive galaxies in HSC surveys by comparing the observed aperture stellar masses
    to the ones predicted by the best--fit \asap{} model. 
    Here, we briefly demonstrate the performances of two \mvir{} estimators here:
    the random forest regressor and the 2--D \mmax{}-\minn{}-\mhalo{} scaling relation. 
    The \texttt{Jupyter} notebook used for \mhalo{} predictions is available here:
    \href{https://github.com/dr-guangtou/asap/blob/master/note/assign_halo_mass.ipynb}{\faGithubAlt}.
    
\begin{figure*}
      \centering 
      \includegraphics[width=\textwidth]{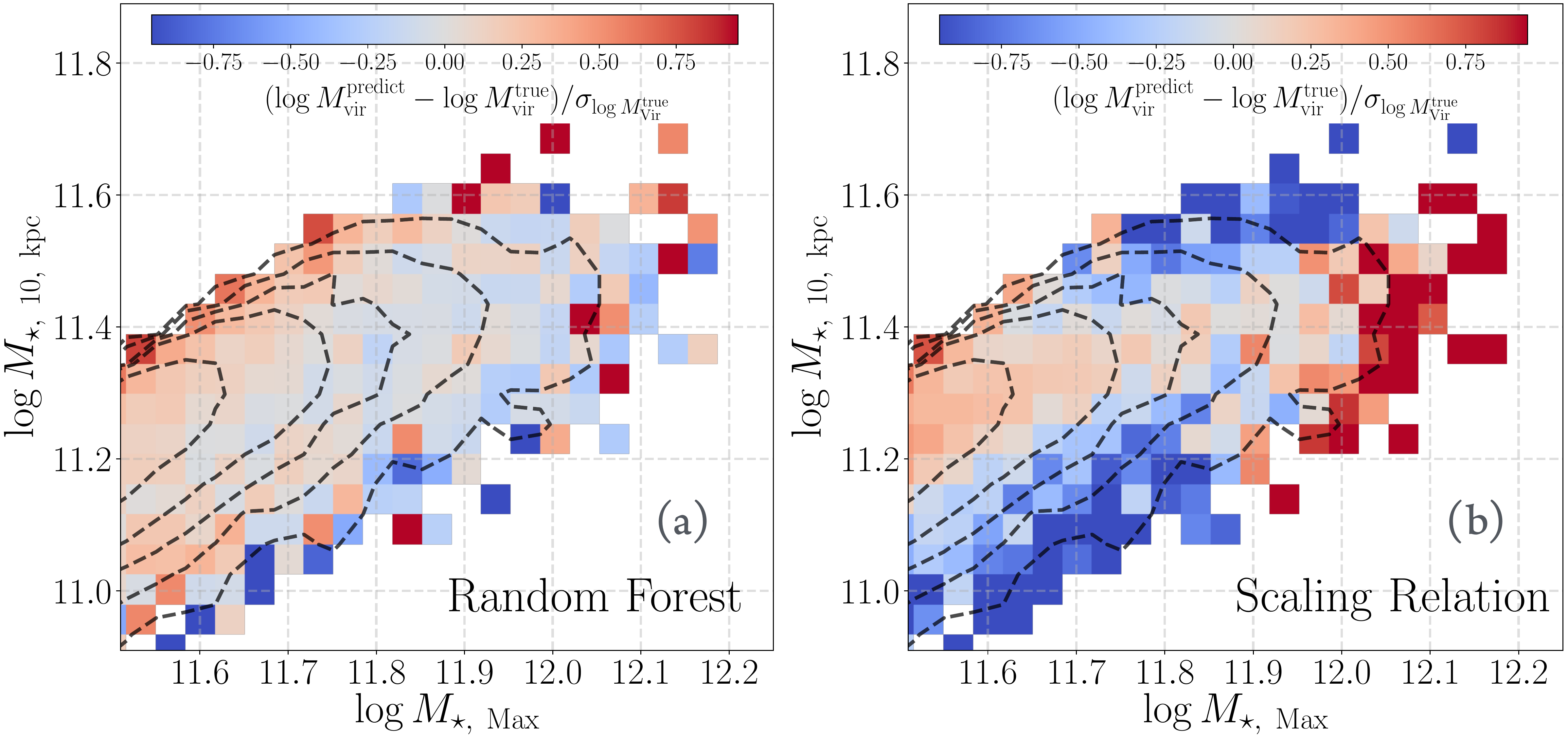}
      \caption{ 
          Evaluation of two different \mvir{} predictors based on the halo mass trend 
          over the aperture mass plane.
          The left panel is for the random forest regressor and the right side is for 
          the \mmax{}--\minn{}--\mvir{} scaling relation. 
          On both figures, the color indicates the relative differences between 
          the predicted \mvir{} and the true values from the \um{} model. 
          Regions occupied by most observed HSC galaxies are highlighted using 
          grey contours. 
          The \texttt{Jupyter} notebook for this figure is available here:
          \href{https://github.com/dr-guangtou/asap/blob/master/note/figB1.ipynb}{\faGithubAlt}
          }
      \label{fig:mvir_predict}
  \end{figure*}

    For the random forest regressor, we use the \texttt{RandomForestRegressor} from the 
    \texttt{sciki-learn} Python package. 
    We choose to use 20 estimators and mean absolute error criteria. 
    We train the random forest regressor using the predicted \mmaxmod{} and \minnmod{}
    of \emph{central} galaxies from the best--fit \asap{} model, then validate it 
    using a realization of the \asap{} model with parameters slightly deviated from the 
    best--fit values. 
    On the left side of Fig~\ref{fig:mvir_predict}, we visualize the performance of this 
    estimator across the aperture stellar mass plane.  
    We choose to use 
    $(\log M_{\rm vir}^{\rm predict} - \log M_{\rm vir}^{\rm true}) / \sigma_{\log M_{\rm vir}^{\rm true}}$
    to test the accuracy of the prediction where $\sigma_{\log M_{\rm vir}^{\rm true}}$ 
    is the scatter of \mvir{} in each 2--D bin of aperture masses. 
    As expected, the random forest regressor can easily capture the detailed 
    trend of \mhalo{} over the 2--D aperture stellar mass plane. 
    
    Meanwhile, we have shown the best--fit 2--D \mmax{}-\minn{}-\mhalo{} scaling relation 
    in \S\ref{ssec:mhalo_tren}.
    We visualize its accuracy on the right side of Fig~\ref{fig:mvir_predict}. 
    As one can see, this simple scaling relation can still capture the main 
    \mhalo{} trend over the regions that are occupied by most HSC galaxies 
    (highlighted by contours). 
    Although the \mhalo{} predicted by this 2--D scaling relation starts to show
    deviations compared to true values at the edges of the aperture stellar mass
    relation, the systematic differences are still comparable to the scatters of \mhalo{}
    in these bins.
    
    In the next paper of this series, we will be looking for more reliable way to 
    predict \mhalo{} based on the stellar mass distributions of massive galaxies and 
    improved version of \asap{} model. 
    We will also directly test these predictions using HSC weak lensing calibrations.

\bsp
\label{lastpage}
\end{document}